\documentclass[12pt,a4paper]{article}
\pdfoutput=1
\usepackage{jheppub}
\usepackage{cite}

\usepackage{caption}
\usepackage{subcaption}
\usepackage{array,arydshln}
\usepackage{braket}
\usepackage{amsmath,amssymb}

\usepackage{here}

	\newcommand{\be}{\begin{equation}}
		\newcommand{\ee}{\end{equation}}
	\newcommand{\bea}{\begin{eqnarray}}
		\newcommand{\eea}{\end{eqnarray}}
	\newcommand{\beas}{\begin{eqnarray*}}
		\newcommand{\eeas}{\end{eqnarray*}}
	
	\def\order{\ensuremath{\mathcal{O}}}

\newcommand{\pho}{F}

\title{Krylov complexity in the IP matrix model I\hspace{-1.2pt}I}

\author[a]{Norihiro Iizuka} 
\author[b]{and Mitsuhiro Nishida}

\affiliation[a]{Department of Physics, Osaka University, Toyonaka, Osaka 560-0043, JAPAN}
\affiliation[b]{Department of Physics, Pohang University of Science and Technology, Pohang 37673, Korea}

\emailAdd{iizuka@phys.sci.osaka-u.ac.jp} 
\emailAdd{nishida124@postech.ac.kr}

\abstract{We continue the analysis of the Krylov complexity in the IP matrix model. In a previous paper, \cite{Iizuka:2023pov}, for a fundamental operator, it was shown that at zero temperature, the Krylov complexity oscillates and does not grow, but in the infinite temperature limit, the Krylov complexity grows exponentially in time as $\sim \exp\left( {\order\left( {\sqrt{t}}\right)} \right)$. We study how the Krylov complexity changes from a zero-temperature oscillation to an infinite-temperature exponential growth. At low temperatures, the spectral density is approximated as collections of infinite Wigner semicircles. We showed that this infinite collection of branch cuts yields linear growth to the Lanczos coefficients and gives exponential growth of the Krylov complexity. Thus the IP model for any nonzero temperature shows exponential growth for the Krylov complexity even though the Green function decays by a power law in time. We also study the Lanczos coefficients and the Krylov complexity in the IOP matrix model taking into account the $1/N^2$ corrections. There, the Lanczos coefficients are constants and the Krylov complexity does not grow exponentially as expected. 
}

\begin{document}

\begin{flushright}
{\small OU-HET-1198}
 \\
\end{flushright}

\maketitle

\section{Introduction}

Krylov complexity is proposed as a new diagnostics for chaos in \cite{Parker:2018yvk}. It is defined for each local operator $\mathcal{O}$ and it captures how that local operator $\mathcal{O}$ keeps changing in a time evolution in the subspace of operator algebra, called Krylov subspace. It is conjectured that in a chaotic system, Krylov complexity grows exponentially in time. The famous example which shows exponential growth is the SYK model \cite{Sachdev:1992fk, Kitaevtalk}, which is known to show chaotic behavior in the low-temperature limit. For other literature, see for examples \cite{Barbon:2019wsy, Avdoshkin:2019trj, Cao:2020zls, Jian:2020qpp, Dymarsky:2019elm, Yates:2020lin, Yates:2020hhj, Rabinovici:2020ryf, Rabinovici:2021qqt, Yates:2021lrt, Yates:2021asz, Dymarsky:2021bjq, PhysRevE.104.034112, Trigueros:2021rwj, Liu:2022god, Fan:2022xaa, Kar:2021nbm, Caputa:2021sib, Balasubramanian:2022tpr, Heveling:2022hth, Adhikari:2022whf, Bhattacharjee:2022qjw, Bhattacharjee:2022vlt, Du:2022ocp,  Banerjee:2022ime, 
Muck:2022xfc, Hornedal:2022pkc, Guo:2022hui, Rabinovici:2022beu, Alishahiha:2022anw, Avdoshkin:2022xuw, Camargo:2022rnt, Kundu:2023hbk, Rabinovici:2023yex, Zhang:2023wtr, Nizami:2023dkf, Hashimoto:2023swv, Nandy:2023brt,  Caputa:2022eye, Caputa:2022yju, Afrasiar:2022efk, Balasubramanian:2022dnj, Erdmenger:2023shk, Bhattacharya:2022gbz, Bhattacharjee:2022lzy, Bhattacharya:2023zqt, Chattopadhyay:2023fob, Pal:2023yik, Patramanis:2023cwz, Bhattacharyya:2023dhp, Camargo:2023eev,  Caputa:2023vyr, Fan:2023ohh, Vasli:2023syq, Bhattacharyya:2023zda, Gautam:2023bcm,Suchsland:2023cmb}.

To understand the nature of the Krylov complexity better, it is desirable to work out what model shows the exponential behavior of the Krylov complexity. 
In the previous paper, \cite{Iizuka:2023pov}, we study the Krylov complexity of the IP matrix model \cite{Iizuka:2008hg} both at zero-temperature limit $(T=0)$ and also at the infinite-temperature limit $(T=\infty)$. 
The IP matrix model is a simple large $N$ matrix model,  
introduced previously as a toy model of the gauge theory dual of an AdS black hole. At finite $N$, this model is an interacting finite number of harmonic oscillators; thus, nothing strange can happen.  
However, only in the large $N$ limit, this model shows key signatures of thermalization and information loss: even though the spectrum is discrete at zero temperature, in the high-temperature limit, its spectrum becomes continuous and gapless and the Green function decays exponentially in time, indicating information loss. 
How the spectrum changes from the zero-temperature discrete one to the infinite temperature continuum and gapless one is quite nontrivial. 

In the previous paper \cite{Iizuka:2023pov}, our focus is both zero-temperature limit and infinite-temperature limit, and we showed that at the infinite temperature limit, the Lanczos coefficients $b_n$ grow linearly in $n$ with logarithmic corrections, which is one of the fastest growth under certain conditions. Krylov complexity then grows exponentially in time as $\sim \exp\left({\order{\left(\sqrt{t}\right) }}\right)$. These show that the IP model at sufficiently high temperatures is chaotic, 
and as far as we are aware, this exponential growth of the Krylov complexity is the first example found in the large $N$ matrix model. 

This paper aims to fill the gap between $T=0$ and $T = \infty$. We study more in detail about the Lanczos coefficients and the Krylov complexity, especially at finite nonzero temperatures. In the temperature range between $T=0$ and $T = \infty$, the spectrum changes from discrete to continuous but dependent on the temperatures the spectrum shows gaps at low temperatures but these gaps close at high temperatures.

The organization of this work is as follows. After a short summary of the results of previous papers  \cite{Iizuka:2023pov} in section \ref{IPreview}, we study the Lanczos coefficients and the Krylov complexity based on toy models that capture the essential features of the IP model at finite temperature in section \ref{mainanalysis}.  Section \ref{mainnumerics} is for numerical analysis of the IP model. We also study the Lanczos coefficients and the Krylov complexity for the IOP model \cite{Iizuka:2008eb} in section \ref{IOPKrylov}. 
We end with conclusions and speculation in section \ref{conclusion}. Appendix \ref{review} is for the basics of Lanczos coefficients and Krylov complexity. 


\section{Review of IP matrix model, Lanczos coefficient and Krylov complexity}\label{IPreview}
We briefly review a spectrum of the IP matrix model, the Lanczos coefficients, and the Krylov complexity in both $T=0$ and $T = \infty$ limit. For more details, see \cite{Iizuka:2023pov}. 

\subsection{The IP matrix model}\label{sec:IP}

IP model contains 
a Hermitian matrix field $X_{ij}(t)$ and a complex vector field $\phi_i(t)$. $X_{ij}(t)$ and $\phi_i(t)$ are harmonic oscillators with masses $m$ and $M$, in the $U(N)$ adjoint and fundamental representations, respectively. 
They obey the conventional quantization condition, 
\begin{align}
[ X_{ij}, \Pi_{kl} ] = i \delta_{il} \delta_{jk} \,, \quad  [\phi_i, \pi_j] = i \delta_{ij} \,.
\end{align} 
The Hamiltonian is 
\begin{eqnarray}
H = \frac{1}{2} {\rm Tr}(\Pi^2) + \frac{m^2}{2} {\rm Tr}(X^2) + M(a^\dagger a + \bar a^\dagger \bar a)  
+ g (a^\dagger X a + \bar a^\dagger X^T \bar a)\ , \label{ham}
\end{eqnarray}
where $a^\dagger_i$ and $a_i$ are creation/annihilation operator for a vector field $\phi_i$, 
\begin{equation}
a_i = \frac{ \pi_i^\dagger  - i M\phi_i}{\sqrt{2M}}\ ,\quad
\bar a_i = \frac{ \pi_i  - i M\phi_i^\dagger}{\sqrt{2M}}\,.
\end{equation}

We consider the following time-ordered retarded Green's function for the fundamental as observable, 
\be\label{tpa}
e^{i M (t-t')} \Big{\langle} \mbox{T} \, a_i(t)\, a_j^\dagger(t') \Big{\rangle}_T := \delta_{ij} G(T, t-t') \,, 
\ee
where $T$ is temperature.  
We always consider the limit $M \gg m > 0$\footnote{We can take the $m\to0$ limit as well. However, then the spectral density is given by a single Wigner semicircle and we will not obtain any interesting large $N$ transition between $T=0$ and $T \neq 0$ \cite{Iizuka:2008hg}. Thus, we focus on $m > 0 $ in this paper.} and $M \gg T$.
With 't Hooft coupling $\lambda := g^2 N$, the Schwinger-Dyson equations for the fundamental in the large $N$ limit becomes 
\be
\label{IPSDeq}
\tilde{G}(\omega) = \tilde{G}_0(\omega) - \lambda \tilde{G}_0(\omega) \tilde{G}(\omega) \int_{-\infty}^{\infty}  \frac{d \omega'}{2 \pi} \tilde{G}(\omega') \tilde{K}(T, \omega- \omega')
\ee
where $\tilde{G}$ is a dressed propagator and $\tilde{G}_0$ is a temperature-independent bare propagator, 
\begin{align}\label{freepropagator}
\tilde{G}_0  = \frac{i}{\omega + i \epsilon} \,.
\end{align} 
$\tilde{K}$ is a thermal propagator for $X_{ij}$, given by
\begin{align} 
\tilde K(T,\omega) = \frac{i}{1 - e^{- m/T}} \left( \frac{1}{\omega^2 - m^2 + i\epsilon}
- \frac{e^{- m/T}}{\omega^2 - m^2 - i\epsilon} \right)  . \label{ktherm}
\end{align}
which is free since the backreaction of $X_{ij}$ on $\phi_i$ is suppressed by $1/N$. 
Using the fact that 
the time-ordered correlator $G(T,t)$ vanishes at $t < 0$,  
by closing the contour in the upper half-plane, we have 
\begin{align}
\hspace{-9mm}\tilde G(T,\omega - m)  - \frac{4}{\nu_T^2}\frac{1}{ \tilde G(T,\omega)} + e^{- m/T} \tilde G(T,\omega + m) 
= \frac{4 i \omega}{\nu_T^2}  \label{sde4} \,,
\end{align}
where 
\begin{align}
\nu_T^2 := \frac{\nu^2}{1- e^{- m/T}} \,, \quad \nu^2 := \frac{2\lambda}{m} \,. 
\end{align}
Since the real part of $\tilde{G}$ is a spectral density, defining 
\begin{align}
{\rm Re}\,\tilde G(T,\omega) =  \pi \pho(\omega) 
\end{align}
$\pho(\omega)$ is the spectral density.  
Then eq.~\eqref{sde4} reduces to 
\begin{equation}
\pho(\omega - m)  - \frac{4}{\nu_T^2 | \tilde G(T,\omega) |^2} {\pho(\omega)} + e^{- m/T}
\pho(\omega + m) = 0 .
\label{real} 
\end{equation}
This recursion relation is the key equation of the model. 
If $F(\omega_0) =0$ at some $\omega = \omega_0$, then $F(\omega_0 \pm m) =0$ as well. Inversely if $F(\omega_0) \neq 0$ at some $\omega = \omega_0$, then $F(\omega_0 \pm m) \neq 0$, and thus, there are infinite amounts of cuts in both positive $\omega$ as well as negative $\omega$. 

The fact that each cut is accompanied by an unbounded series of additional cuts in both positive and negative $\omega$ directions can be seen by employing proof by contradiction as follows; Suppose the branch cuts continue only up to $\omega = \omega_0$. In other words, suppose that 
\begin{align}
\pho(\omega_0) \neq 0  \,, \quad  \mbox{but} \quad \pho(\omega_0 + m) = 0  \, \quad \mbox{at some $\omega_0$}
\end{align}  
Then setting $\omega = \omega_0 + m $ in \eqref{real},  we obtain 
\begin{align}
\pho(\omega_0 ) + e^{- m/T}
\pho(\omega_0 + 2 m) = 0 \,.
\end{align}
However, due to the positivity of $\pho(\omega_0 )$, this immediately implies $\pho(\omega_0 )  = 0$, which is in contradiction. Similarly one can show that there is no bound on the lower $\omega$ direction as well. Thus, each cut is always accompanied by an unbounded series of additional cuts, and this fact plays a key role in the Lanczos coefficients and the Kryloc complexity analysis. 
Note that the structure that there is an infinite amount of $m$-translated cut exists only in nonzero temperature.  $T=0$ is special since there are poles and $| \tilde G(T,\omega) |$ diverges. But in nonzero $T > 0$, the pole disappears.

\subsection{The spectral density}

The IP model spectrum for $m \neq 0$ changes drastically from collections of discrete poles at $T=0$ to continuum and gapless spectrum at $T = \infty$. See \cite{Iizuka:2008hg, Iizuka:2023pov} for detail. Here we comment $T=0$ and $T = \infty$ results. 

\subsubsection{Zero temperature: $T=0$}

In the zero temperature case,  
one can solve this equation analytically by mapping the equation to the Bessel recursion relation based on the canonical calculation \cite{Iizuka:2008hg},  
\begin{align}\label{GzeroT}
\tilde G(\omega)  
= -\frac{2i}{\nu}\frac{ J_{-\omega/m}(\nu/m)}{J_{-1-\omega/m}(\nu/m)}\ .
\end{align}
where $J$ is a Bessel function  
The spectrum is determined by the pole of $\tilde G(\omega) $ which is discrete for nonzero $m > 0$. There are infinite poles, which are determined by the zeros of the denominator.

\subsubsection{Infinite temperature limit: $T =\infty$}

In the infinite temperature limit, $T = \infty$, the recursion relation \eqref{real} becomes 
\begin{equation}
\pho(\omega - m)  - \frac{4}{\nu_T^2 | \tilde G(T,\omega) |^2} {\pho(\omega)} +
\pho(\omega + m) = 0 .
\label{realTinfty} 
\end{equation}
This recursion relation is symmetric between $\omega \to \infty$ to $\omega \to - \infty$.  
Furthermore, there are $m$ shifted structure: if $F(\omega_0) \neq 0$ at some $\omega = \omega_0$, then $F(\omega_0 \pm m) \neq 0$. In fact, as we increase the temperature, the spectrum change from collections of poles to collections of branch cuts, and finally these cuts merge and the spectrum becomes gapless and continuous \cite{Iizuka:2008hg, Iizuka:2023pov}. 
In large $\omega$, $F(\omega)$ decays exponentially as $\pho(\omega) \sim |\omega|^{-|\omega|}$. In fact, the asymptotic solution can be obtained as \cite{Iizuka:2023pov}
\begin{align}
\pho(\omega) 
\sim&\exp\left[-\frac{2\vert\omega\vert}{m}\log\left(\frac{2\vert\omega\vert}{\nu_T}\right)\right] \;\;\; (\omega\to \pm\infty).\label{sed+}
\end{align}
This exponential decay of the $F(\omega)$ (with $\log \omega$ correction) leads to the Lanczos coefficients $b_n$ growing asymptotically linear in $n$ (with $\log n$ correction) at large $n$.

\subsection{Lanczos coefficients and Krylov complexity in the IP model}\label{sec:LanczosIP} 

We summarize the basic aspects of Lanczos coefficients and Krylov complexity in the appendix \ref{review}. 
Since our observable is given by $G(T, t-t')$ in eq.~\eqref{tpa}, let us evaluate the Lanczos coefficients of the IP model in the large $N$ limit associated with $\hat{\mathcal{O}}=\hat{a}^\dagger_j$ and its two-point function 
\begin{align}
C(t;\beta):=e^{i M t}(\hat{a}_j^\dagger(t)\vert \hat{a}_j^\dagger(0))_\beta \,.
\end{align}
Here $C(t;\beta)$ is {\it not} a time-ordered correlator, and thus different from $G(t)$ given by eq.~\eqref{tpa}.

In fact, the Fourier transformation $f(\omega)$ of $C(t;\beta)$ can be expressed by $\pho(\omega)$. Since 
\begin{align}\label{timereversal}
C^*(t;\beta)
&=e^{-i M t}(\hat{a}_j^\dagger(-t)\vert \hat{a}_j^\dagger(0))_\beta=C(-t;\beta)\,,
\end{align}
we obtain 
\begin{align} 
f(\omega) 
&=\tilde G(T,\omega)+\tilde G^*(T,\omega)=2\pi \pho(\omega) \,. \label{Ffrela}
\end{align}
The key point is that $G(T,t)$ vanishes at $t<0$ and $G(T,t)=C(t;\beta)$ at $t>0$. At least numerically, we can compute $f(\omega)=2\pi \pho(\omega)$ by solving the recursion relation (\ref{real}).

\subsubsection{Zero temperature: $T=0$}

At zero temperature, let $\vert v\rangle$ be the free ground state. Then, consider excited states $\vert j, n\rangle$ with a single excitation by $\hat{a}^\dagger_i$ and $n$-excitations by $\hat{A}^\dagger_{ij}$ such as
\begin{align}\label{IPbasiszerotemperature}
\vert j, n\rangle:=i^{-n}N^{-n/2}\hat{a}_i^\dagger (\hat{A}^{\dagger n})_{ij}\vert v\rangle.
\end{align}
In the large $N$ limit, the states $\vert j, n\rangle$ span an orthonormal basis for the two-point function. 
Moreover, in the large $N$ limit, we obtain \cite{Iizuka:2008hg}
\begin{align}
(H-M)\vert j,n\rangle=mn \vert j,n\rangle+\frac{\nu}{2}\vert j,n-1\rangle+\frac{\nu}{2}\vert j,n+1\rangle,
\end{align}
Therefore, with $\mathcal{L}=H-M$, the Krylov basis for $\hat{\mathcal{O}}=\hat{a}^\dagger_j$ is $\vert \hat{\mathcal{O}}_n)=\vert j, n\rangle$ given by eq.~\eqref{IPbasiszerotemperature}. Then the Lanczos coefficients are given by
\begin{align}\label{LczeroT}
a_n=mn, \;\;\; b_n=\frac{\nu}{2}.
\end{align}
and $b_n$ in the IP model does not depend on $n$ because we determine the normalization of (\ref{IPbasiszerotemperature}) in the large $N$ limit.

Given the Lanczos coefficients as eq.~\eqref{LczeroT}, we perform numerical calculations of $K(t)$. 
The resultant Krylov complexity $K(t)$ oscillates due to nonzero $a_n$ and does not grow at late times \cite{Iizuka:2023pov}.

\subsubsection{Infinite temperature limit: $T = \infty$}

The asymptotic behavior of the spectral density \eqref{sed+} determines the Lanczos coefficients of the IP model at the infinite temperature limit. 
The spectral density $F(\omega)$ is a positive and smooth function everywhere \cite{Iizuka:2008hg} 
and $F(\omega)$ is an even function with respect to $\omega$. Thus $a_n=0$. 
Furthermore the exponential suppression in \eqref{sed+}
is known as the slowest decay for the situation where $C(t)$ is analytic in the entire complex time plane \cite{Parker:2018yvk, Avdoshkin:2019trj}. 
From \eqref{sed+}, the asymptotic behavior of $b_n$ of the IP model in the infinite temperature limit can be determined as
\begin{align}
b_n\sim \frac{m\pi n}{4W(2m\pi n/\nu_T)}\sim \frac{m\pi n}{4\log n}  \;\;\; (n\to\infty).\label{abbnIP}
\end{align}
Here $W(n)$ is defined by 
$z=W(z e^z)$,
 called the Lambert W function.  

Given the asymptotic behavior of $b_n$ 
with $a_n=0$. 
Then, the late-time behavior of $K(t)$ is given as 
\begin{align}
K(t)\sim e^{\sqrt{m\pi t}} = \sum_n \frac{\left( m \pi t\right)^{\frac{n}{2}}}{n!}\,.\label{Ktroott}
\end{align}
This growth behavior is slower than the exponential growth $e^{\order{\left( t \right)}}$, but it is faster than any power low growth behavior for integral systems. The growth of $b_n$ as $b_n \propto n/{\log n}$ strongly suggests that the IP model in the infinite temperature limit is chaotic.

\section{Lanczos coefficient and Krylov complexity for nonzero $T$}
\label{mainanalysis}

\begin{figure}[H]
\vspace{-8mm}
\centering
     \begin{subfigure}[b]{0.35\textwidth}
         \centering
         \includegraphics[width=\textwidth]{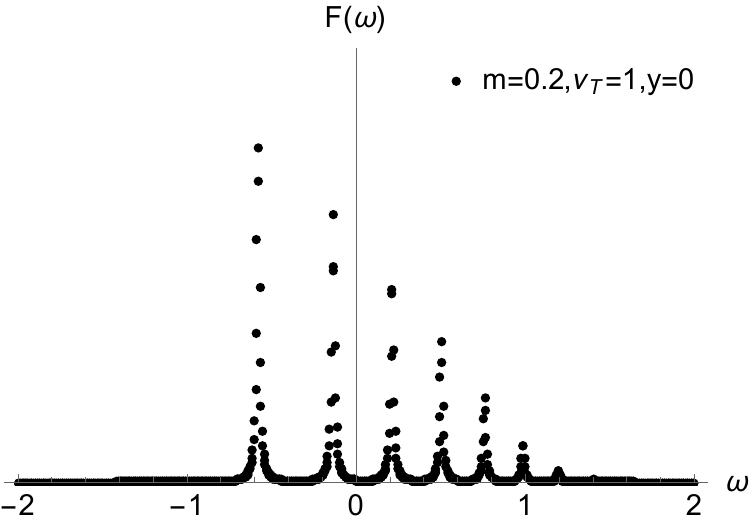}
     \end{subfigure}
      \quad\quad\quad
          \begin{subfigure}[b]{0.35\textwidth}
         \centering
         \includegraphics[width=\textwidth]{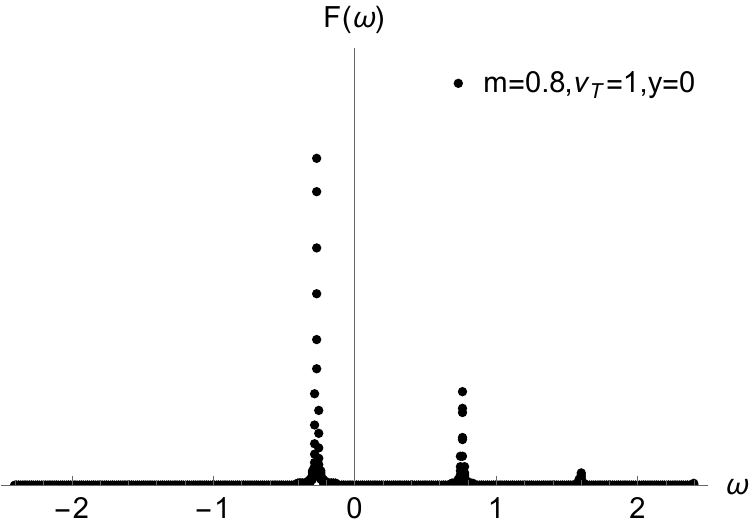}
     \end{subfigure}
     \begin{subfigure}[b]{0.35\textwidth}
         \centering
         \includegraphics[width=\textwidth]{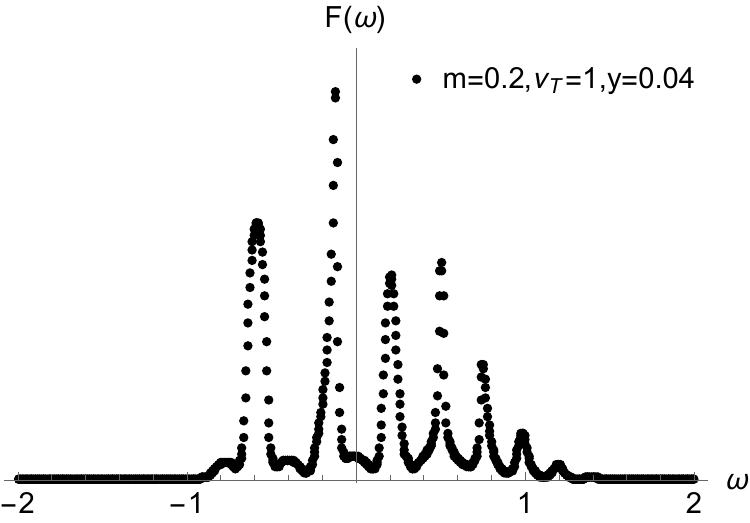}
     \end{subfigure}
      \quad\quad\quad
        \begin{subfigure}[b]{0.35\textwidth}
         \centering
         \includegraphics[width=\textwidth]{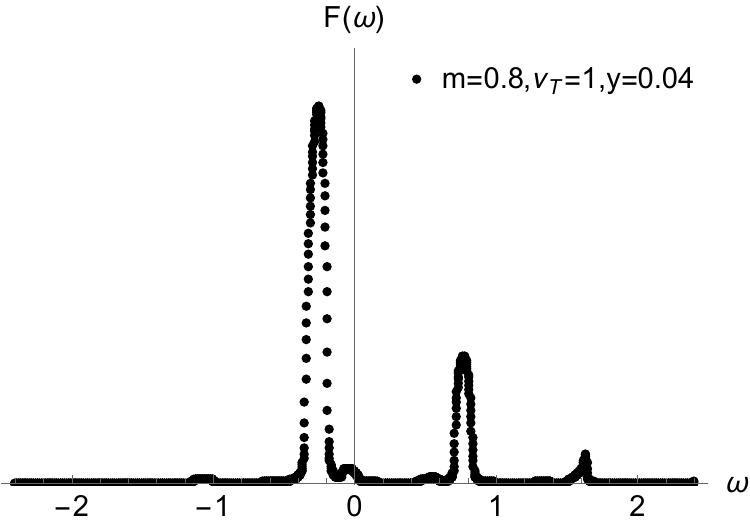}
     \end{subfigure}
         \begin{subfigure}[b]{0.35\textwidth}
         \centering
         \includegraphics[width=\textwidth]{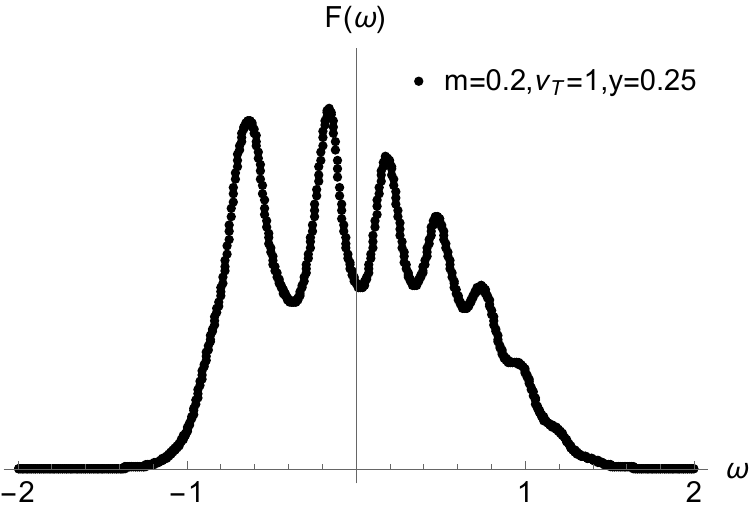}
     \end{subfigure}
      \quad\quad\quad
       \begin{subfigure}[b]{0.35\textwidth}
         \centering
         \includegraphics[width=\textwidth]{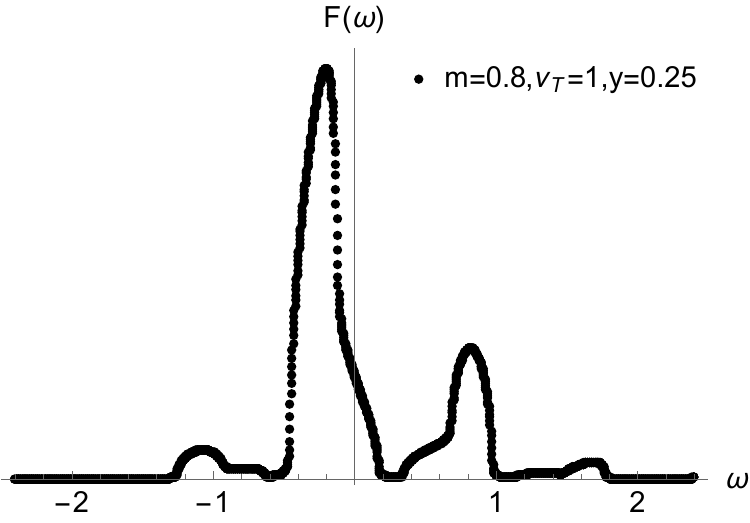}
     \end{subfigure}
            \begin{subfigure}[b]{0.35\textwidth}
         \centering
         \includegraphics[width=\textwidth]{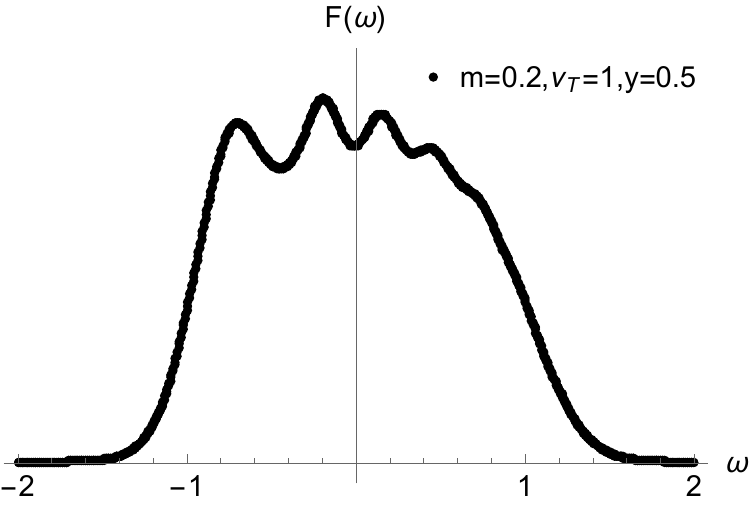}
     \end{subfigure}
      \quad\quad\quad
       \begin{subfigure}[b]{0.35\textwidth}
         \centering
         \includegraphics[width=\textwidth]{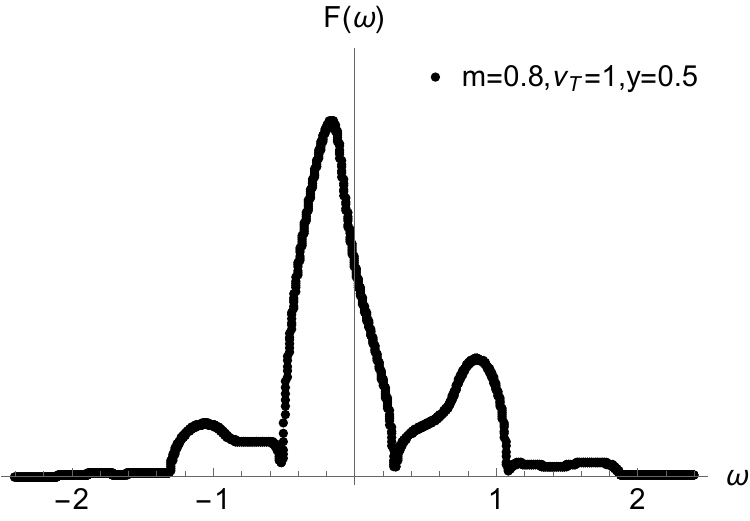}
     \end{subfigure}
       \begin{subfigure}[b]{0.35\textwidth}
         \centering
         \includegraphics[width=\textwidth]{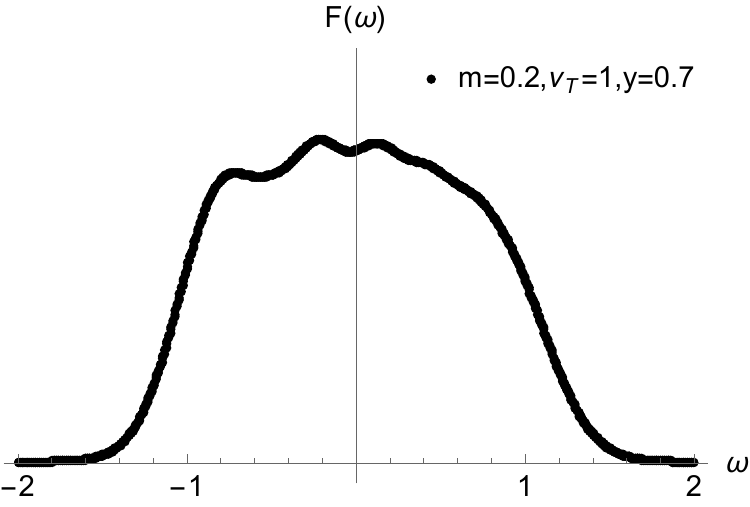}
     \end{subfigure}
      \quad\quad\quad
       \begin{subfigure}[b]{0.35\textwidth}
         \centering
         \includegraphics[width=\textwidth]{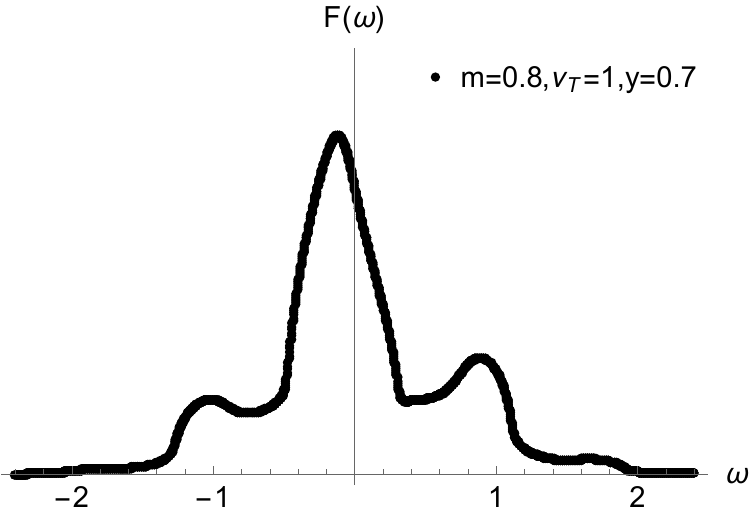}
     \end{subfigure}
              \begin{subfigure}[b]{0.35\textwidth}
         \centering
         \includegraphics[width=\textwidth]{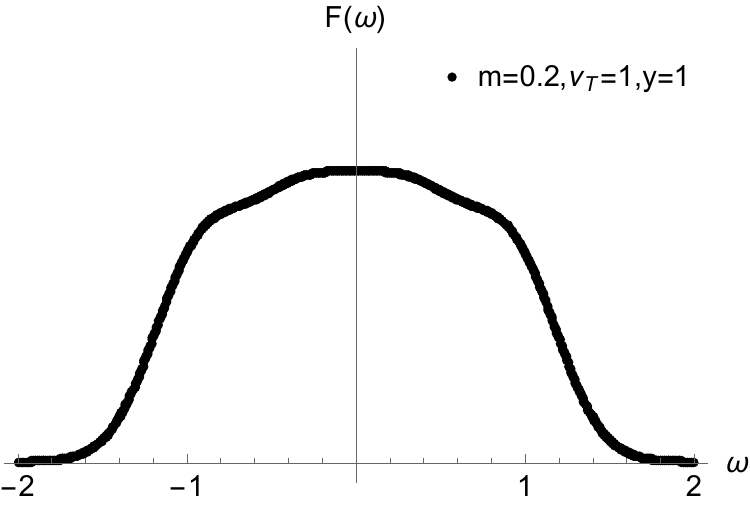}
              \end{subfigure}
         \quad\quad\quad
          \begin{subfigure}[b]{0.35\textwidth}
         \centering
         \includegraphics[width=\textwidth]{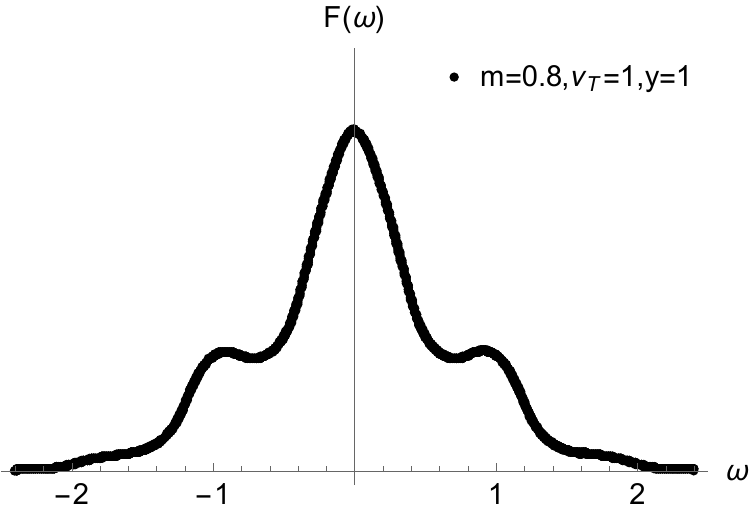}
     \end{subfigure}
             \caption{Numerical plots of $F(\omega)$ for $\nu_T = 1$, $m=0.2$ and $m=0.8$ at various temperatures from $T=0$ to $T=\infty$ where $y=e^{-m/T}$. This is the same figure as Figure 4 in  \cite{Iizuka:2023pov}.}
        \label{fig:F(w)}
\end{figure}

Given the analysis of the Krylov complexity at both $T=0$ and $T=\infty$ in the IP model, in this paper, we focus on the Lanczos coefficient and Krylov complexity with nonzero mass $m > 0$ at finite and nonzero temperatures, {\it i.e.,} $0 < T < \infty$. In Figure \ref{fig:F(w)}, the spectrum density at various temperatures are shown. For Figure \ref{fig:F(w)}, we set $\epsilon$ in (\ref{freepropagator}) as $\epsilon=0.01$ for $m=0.2$, and $\epsilon=0.005$ for $m=0.8$. We also consider a similar shift of $\omega$ in the propagator at $T=0$ as $\tilde{G}(0,\omega+i\epsilon)$. Several comments are in order.

\begin{itemize}
\item At $T=0$, the spectrum is a collection of discrete poles (delta functions). 
However as we increase the temperature, these poles become short branch cuts. 
\item At nonzero but low temperatures $T < T_c$, there is an infinite short branch cut and their positions are related by multiples of $m$.  Here, $T_c$ is the critical temperature at which the spectrum becomes gapless. 
\item The critical temperature in Figure \ref{fig:F(w)} is $y=e^{-m/T_c}\sim0.1$  for $m=0.2$, and $y=e^{-m/T_c}\sim0.3$  for $m=0.8$. By taking into account for errors in numerical calculations, we attribute the existence of the periodic energy gap to a region where $F(\omega)\lesssim10 \, \epsilon$, near $\omega=0$. 
\item As we increase the temperature, these branch cuts become longer and at some point, $T=T_c$, they merge. 
\item At high temperatures  $T > T_c$, the cuts merge completely and the spectrum becomes gapless. 
\item In the infinite temperature limit $T = \infty$, the spectrum becomes gapless and symmetric under $\omega \leftrightarrow - \omega$. 
\item As $T$ increases, the behavior of a two-point function $C(t;\beta)=\int_{-\infty}^{\infty}\frac{d\omega}{2\pi} f(\omega)e^{-i\omega t}$ changes as follows \cite{Iizuka:2008hg}. At $T=0$, $C(t;\beta)$ oscillates and does not decay due to delta functions in the discrete spectrum. At nonzero temperature $0<T$, the spectrum has no poles on the real axis of $\omega$, and $C(t;\beta)$ should decay asymptotically.  At nonzero low temperature $0<T < T_c$, $C(t;\beta)$ decays by a power law since the gapped spectrum is not smooth on the real axis. At high temperature $T_c\ll T$, the spectrum is smooth on the real axis, and $C(t;\beta)$ decays exponentially.
\end{itemize}

\clearpage

\subsection{Low temperature $T_c > T>0$ gapped spectrum and models for spectral density} 
Let's first try to understand the low-temperature behavior where the spectrum is continuous with gaps. 
See $y = e^{-m/T} = 0.04$ in Figure \ref{fig:F(w)}, where there are infinite\footnote{Although there are infinite cuts, since their magnitudes decay exponentially numerically at large $|\omega|$, in Figure \ref{fig:F(w)}, it appears as if there are only finite cuts.} cuts but the length of cuts is short enough that there are gaps between each cut. 

The gapped structure is determined by the key equation eq.~\eqref{real}. There is a $m$-shifted structure of the infinite cuts in $\omega$, {\it i.e.,} if $F(\omega_0) \neq 0$ at some $\omega = \omega_0$, then $F(\omega_0 \pm m) \neq 0$ as well. Therefore if there is a gap at some $\omega$, then that gap exists after $\pm m$ shift in $\omega$. 
Furthermore, 
In the large $\omega \to \infty$, eq.~\eqref{real} allows 
\begin{align} 
f(\omega) = 2 \pi \pho(\omega) \sim  |\omega|^{\order{(\omega})} \,.
\end{align} 
Thus, the key properties in nonzero low-temperature spectral density are that 
\begin{enumerate}
\item There are an infinite amount of gaps and cuts related by $\omega \to \omega \pm m$.  
\item The magnitudes of the spectrum decay exponentially in large $|\omega|$ asymptotically. 
\end{enumerate}

To understand essential properties for such spectral density, let us introduce the following toy model whose spectral density consists of an infinite number of Wigner semicircles with widths $2 \ell$ centered at $\omega = m j$ as follows 
\begin{align}
f(\omega)  = \mathcal{N}\sum_{j=-N_W}^{N_W} f_j(\omega)\,, \label{SpectraSemicircles} \quad \mbox{where} \quad 
f_j(\omega) =\text{Re}\left[\frac{A_j}{\ell^2}\sqrt{\ell^2-(\omega-\omega_j)^2}\right] \,, 
\end{align}
where  $\omega_j$ are centers of the semi-circles and $A_j$ represent amplitudes of the semicircles. This spectral density consists of $2 N_W +1$ numbers of semicircles.  As we have seen in the previous section, the IP model gives $N_W \to \infty$ but for a moment let us keep $N_W$ as a parameter. 

\begin{figure}
         \centering
         \includegraphics[width=0.4\textwidth]{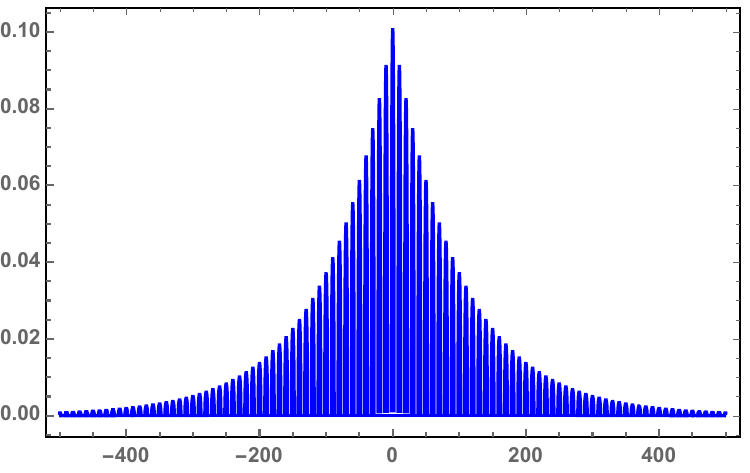}
       \put(5,5){$\omega$}
    \put(-175,115){$f(\omega)$}
       \caption{Spectrum $f(\omega)$ (\ref{SpectraSemicircles}) consisting of 101 Wigner semicircles with (\ref{SpectrumEx1}).}\label{fig:SpectrumWignerSemicircles}
\end{figure}

From the recursion relation \eqref{real}, the gaps in the spectrum occur periodically, and thus we set  
\begin{align}
 \omega_j = m j  \,, \quad \quad A_j=e^{-\vert\omega_j\vert/\Omega} \,, \label{SpectraSemicircles2} 
\end{align}
such that $m$ shifted structure is maintained and asymptotically the magnitudes of the spectrum decay exponentially in $\omega$.  
$\mathcal{N}$ is a normalization constant such that
\begin{align}
\int_{-\infty}^{\infty}\frac{d\omega}{2\pi}\,f(\omega) = \int_{-\infty}^{\infty} {d\omega}\,\pho(\omega) =1.
\end{align}
Note that $\ell$ controls the length of the semi-circles, and in the limit $\ell \to 0$, each Wigner semi-circle becomes a delta function (a pole). For simplicity, we choose $f(\omega)$ to be an even function in $\omega$. In this section, we mainly study the Lanczos coefficients and the Krylov complexity associated with this model. In Figure \ref{fig:SpectrumWignerSemicircles}, we plot $f(\omega)$ \eqref{SpectraSemicircles} with $N_W=50$, where $f(\omega)$ consists of 101 Wigner semicircles with the following parameters
\begin{align}\label{SpectrumEx1}
\ell =2 \,, \;\;\; \omega_j  =10\,j \,\,\, (\Leftrightarrow m=10) \,, \;\;\; A_j=e^{-\vert \omega_j\vert/100}  \,\,\,(\Leftrightarrow \Omega=100) \,.
\end{align}

Suppose the spectral density is given by eq.~\eqref{SpectraSemicircles}, then Fourier transforming it, 
a two-point function $C(t)$ is obtained as 
\begin{align}
C(t)  :&=\int^\infty_{-\infty} \frac{d \omega}{2\pi}e^{-i\omega t}f(\omega)
=\mathcal{N}\sum_{j=-N_W}^{N_W} C_j(t),\\
C_j(t):&= \int^\infty_{-\infty} \frac{d \omega}{2\pi}e^{-i\omega t}f_j(\omega)
=\frac{A_j}{2\ell t}J_1(\ell t)e^{-i \omega_j t}.\label{Ci(t)}
\end{align}
where $J_1$ is a Bessel function of the first kind. 
Since the asymptotic behavior of $J_1(\ell t)$ is
\begin{align}
J_1(\ell t)\sim \sqrt{\frac{2}{\pi \ell t}}\cos \left(\ell t -\frac{3\pi}{4}\right) \;\;\; (\vert t\vert \to\infty),
\end{align}
late-time decay of $C_j(t)$ is the power-law decay with oscillation. When the parameters of $f_j(\omega)$ are given by \eqref{SpectraSemicircles2}, 
$C(t)$ can be computed as
\begin{align}\label{C(t)finiteWigner}
C(t) = \frac{\mathcal{N}J_1(\ell t)}{2\ell t}\sum_{j=-N_W}^{N_W} A_j e^{-i \omega_j t} \,.
\end{align}
Especially when $N_W=\infty$, {\it i.e.,} for the spectral density of an infinite number of Wigner semicircles, $C(t)$ becomes 
\begin{align}\label{C(t)InfiniteWigner}
C(t) =\frac{\mathcal{N}J_1(\ell t)}{2\ell t}\frac{\sinh (m/\Omega)}{\cosh (m/\Omega)-\cos(m t)} \propto \frac{1}{t^{3/2}} \quad \mbox{in $t \to \infty$}\,.
\end{align}

Therefore, late-time decay of $C(t)$ for an infinite number of Wigner semicircles is also the power-law decay with oscillation. We note that $C(t)$ has poles at $t=\pm i/\Omega$, which do not exist in the case of a finite number of Wigner semicircles. The presence of the pole in imaginary time $t=\pm i/\Omega$ with nonzero $\ell$ immediately leads to the conclusion that the Krylov complexity grows exponentially in time, which we will discuss more next.

Thus from the simple model whose spectral density is given by eq.~\eqref{SpectraSemicircles}, \eqref{SpectraSemicircles2}, we learn the following lessons for the IP model at low temperatures:
\begin{itemize}
\item  Suppose the spectral density is made up of {\it finite} cuts, and between each cut, there is a gap. Each cut can be approximated by a Wigner semi-circle. Then even though the spectrum is continuous, its Fourier transformation gives only power-law decay in time.  
\item Even if the spectral density is made up of an {\it infinite} number of cuts, where the magnitude of each Wigner decays exponentially asymptotically in large $\omega$, the resultant two-point function still decays by the power-law in time, not exponentially in time. 
\item However in the case where the number of cuts is infinity, the Krylov complexity grows exponentially in time. It is crucial that the spectrum is continuous, {\it i.e.,} the width of semicircles $\ell \neq 0$, and there are infinite cuts. 
\item For the case where the number of cuts are finite, the exponential growth of the Krylov complexity saturates at finite time. 
\item These suggest that for the exponential growth of the Krylov complexity, it is crucial if the spectrum is continuous or not, and furthermore if it has an upper bound in $\omega$ or not. We will see more in detail soon. 
\end{itemize}

Before we continue the analysis of the Krylov complexity, we consider a discrete limit $ \ell \to0$. In that limit, (\ref{C(t)InfiniteWigner}) becomes
\begin{align}\label{C(t)Discrete}
C(t) =\frac{\mathcal{N}}{4}\frac{\sinh (m/\Omega)}{\cosh (m/\Omega)-\cos(m t)},
\end{align}
which is a periodic function of $t$, and Krylov complexity associated to (\ref{C(t)Discrete}) for a discrete spectrum is also periodic with respect to $t$.  Eq.~\eqref{C(t)InfiniteWigner} and \eqref{C(t)Discrete} are one of the punchlines for the low-temperature behavior in the IP model. 

\subsubsection{Lanczos coefficients and Krylov complexity}
We would like to understand the Lanczos coefficient of the spectral density which consists of an infinite number of Wigner semicircles as eq.~\eqref{SpectraSemicircles}. For simplicity, we consider $f(\omega)=f(-\omega)$, which lead $a_n=0$. As we mentioned, we choose $A_j$ decays exponentially with respect to $\vert\omega_j\vert$  as eq.~\eqref{SpectraSemicircles2}, since the spectrum of the IP model decays exponentially at large $\vert \omega\vert$ including logarithmic correction.

Although the IP model forces the spectral density to have an infinite number of cuts, {\it i.e.,} $N_W \to \infty$, 
let us examine the finite $N_W$ {\it i.e.,} the number of Wigner semicircles is finite. By increasing the number of cuts, $N_W$, we would like to see how the Lanczos coefficients change.

\begin{figure}[t]
\centering
     \begin{subfigure}[b]{0.4\textwidth}
         \centering
         \includegraphics[width=\textwidth]{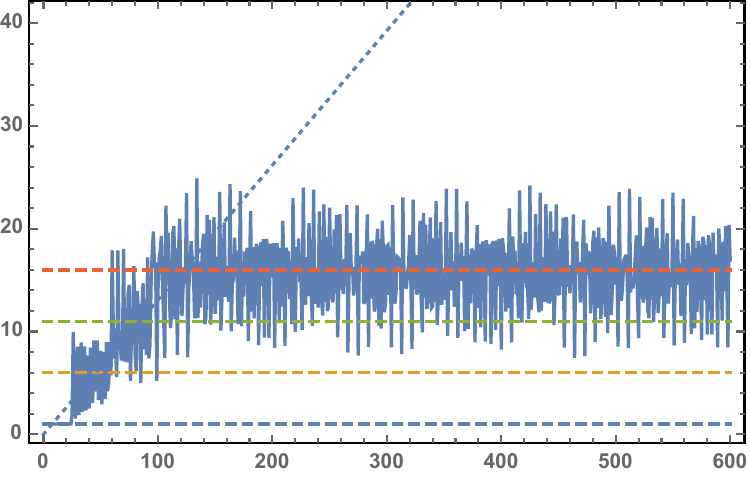}
       \put(5,0){$n$}
    \put(-175,115){$b_n$}
       \caption{$b_n$ for $N_W=3$}\label{fig:bnWignereven(a)}
     \end{subfigure}
      \quad\quad\quad
     \begin{subfigure}[b]{0.4\textwidth}
         \centering
         \includegraphics[width=\textwidth]{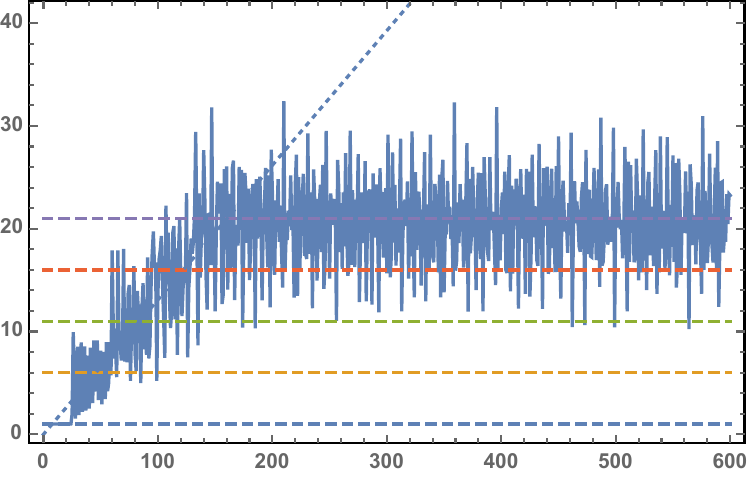}
          \put(5,0){$n$}
     \put(-175,115){$b_n$}
         \caption{$b_n$ for $N_W=4$}\label{fig:bnWignereven(b)}
     \end{subfigure}
             \caption{Lanczos coefficients $b_n$ 
             for the spectral density given by eq.~\eqref{SpectraSemicircles}, \eqref{SpectraSemicircles2} with $\ell =2$, $m=10$, $\Omega=1/12$, (a) $N_W =3$ and (b) $N_W =4$. 
             We connect numerical plots of $b_n$ to make the fluctuation of $b_n$ evident. Horizontal dashed lines are $b_n=\vert\omega_j\vert/2+\ell/2 = m j/2 + \ell/2$. We also plot a dotted line $b_n= \Omega \pi n/2= {\pi n}/{24}$ to compare with the linear growth of $b_n$ where $n$ is small.}
        \label{fig:bnWignereven}
\end{figure}

Figure \ref{fig:bnWignereven} shows Lanczos coefficient $b_n$ for the spectral density given by eq.~\eqref{SpectraSemicircles}, \eqref{SpectraSemicircles2} with $\ell =2$, $m=10$, $\Omega=1/12$, (a) $N_W=3$ (seven semicircles) and (b) $N_W=4$  (nine semicircles) respectively.  

Let us comment on properties of $b_n$ that can be read from Figure \ref{fig:bnWignereven}. 
\begin{itemize}
\item
Around $n=1$, the Lanczos coefficient is a constant $b_n=\ell/2$. This is because $b_n$ around $n=1$ is determined from the single semicircle $f_{j=0}(\omega)$ around $\omega=0$.
\item 
As $n$ increases, the other semicircles begin to contribute to $b_n$. Since the amplitude of semicircles $A_j=e^{-\vert \omega_j\vert/\Omega}$ decays exponentially with respect to $\vert\omega_j\vert$ with $\Omega = 1/12$, $b_n$ increases linearly {\it on average} as $b_n \sim \Omega \pi n/2 = {\pi n}/{24}$. More precisely, due to the existence of gaps in the spectra, $b_n$ increases in a staircase pattern with fluctuation. In our numerical computations, the fluctuation does not grow much as $n$ increases.
\item 
In Figure \ref{fig:bnWignereven}, we consider only a small number of Wigner semicircles in the spectral density, and thus there exists $\omega_{max}$ such that $f(\omega)=0$ for $\vert\omega\vert>\omega_{max} \sim m N_W$. At large $n$, $b_n$ saturates as $b_n\sim \omega_{max}/2$ with fluctuation due to the gaps.
\end{itemize}

\begin{figure}[t]
\centering
         \centering
         \includegraphics[width=10cm]{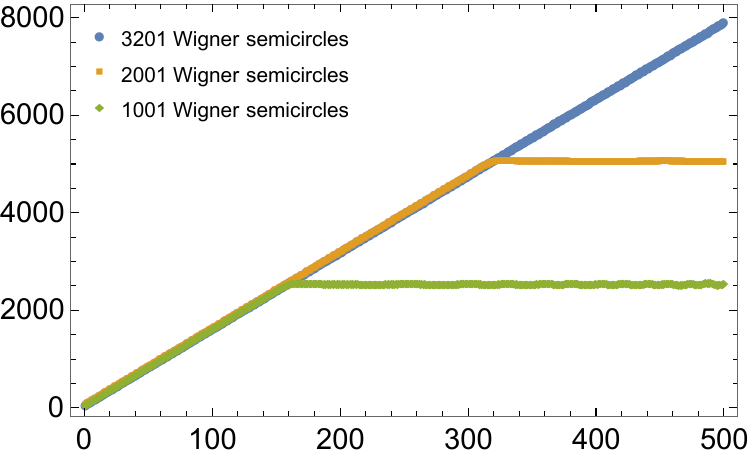}
       \put(5,0){$n$}
    \put(-265,185){$b_n$}
             \caption{Lanczos coefficients $b_n$ with $2N_W + 1$ Wigner semicircles. Here $\ell=2$, $m=10$, $\Omega = 10$, and $N_W = 500$ (green), $1000$ (orange), $1600$ (blue).}
        \label{fig:bnWignereven2}
\end{figure}

Let us consider further examples with a large number of Wigner semicircles, $N_W$. 
Figure \ref{fig:bnWignereven2} shows Lanczos coefficient $b_n$ for the spectral density given by eq.~\eqref{SpectraSemicircles}, \eqref{SpectraSemicircles2} with $\ell =2$, $m=10$, $\Omega=10$,  $N_W=500, 1000, 1600$.

\vspace{-2mm}
\begin{itemize}
\item At large $n$, $b_n$ in Figure \ref{fig:bnWignereven2} saturates as $b_n\sim \omega_{max}/2= m N_W/2= 2500$ for $N_W =500$ and as $b_n\sim \omega_{max}/2 = m N_W/2=5000$ for $N_W = 1000$. The fluctuation of $b_n$ 
in Figures \ref{fig:bnWignereven2} is small compared to the linear growth of $b_n$. The slope of linear growth is $b_n\sim  \pi \Omega n/2 = 5\pi n $.
\item
Comparing Figures \ref{fig:bnWignereven} and \ref{fig:bnWignereven2}, one can check that the linear growth rate of $b_n$ becomes smaller as $\Omega$ decreases.
\item
In the limit that a number of the semicircles is infinite, $N_W \to \infty$, the value of $\omega_{max}$ becomes infinite, and $b_n$ would increase without the saturation even at large $n$ if the amplitude $A_j$ decays with respect to $\omega_j$. 
\end{itemize}

\begin{figure}[t]
\centering
     \begin{subfigure}[b]{0.45\textwidth}
         \centering
         \includegraphics[width=\textwidth]{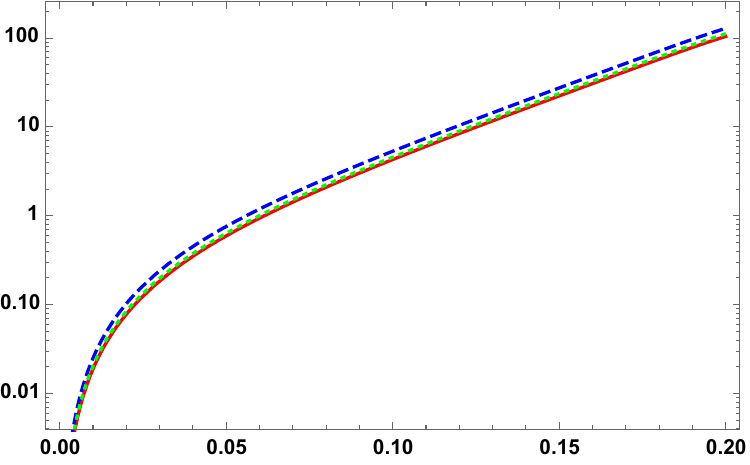}
       \put(5,0){$t$}
    \put(-190,125){$K(t)$}
       \caption{Krylov complexity for $\ell = 2$, $m=10$, $N_W = 1600$, $\Omega=10$.}\label{fig:KCWignereven2(a)}
     \end{subfigure}
      \quad\quad\quad
     \begin{subfigure}[b]{0.45\textwidth}
         \centering
         \includegraphics[width=\textwidth]{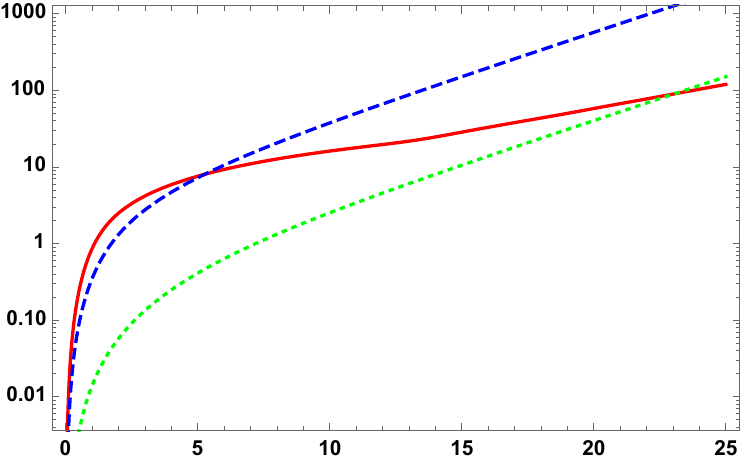}
          \put(5,0){$t$}
     \put(-190,125){$K(t)$}
         \caption{Krylov complexity for $\ell = 2$, $m=10$, $N_W = 1600$, $\Omega=1/12$.}\label{fig:KCWignereven2(b)}
     \end{subfigure}
                  \caption{Log plots of Krylov complexities $K(t)$. Red solid curves are $K(t)$ for the spectral density given by eq.~\eqref{SpectraSemicircles}, \eqref{SpectraSemicircles2} with $\ell =2$, $m=10$, $N_W =1600$ for (a) $\Omega =10$ and (b) $\Omega =1/12$. Blue dashed curves are $K(t)$ computed from the linear fitting of $b_n$ {\it excluding the fluctuation by hand}. Green dotted curves are $K(t)$ for the power spectrum (\ref{GaplessSpectrum}) with $\Omega=10$ in Figure \ref{fig:KCWignereven2(a)} and with $\Omega=1/12$ in Figure \ref{fig:KCWignereven2(b)}.}
        \label{fig:KCWignereven2}
\end{figure}

Once $b_n$ is obtained, we can calculate Krylov complexity numerically. Figure \ref{fig:KCWignereven2} shows log plots of the Krylov complexity $K(t)$ (red solid curves) computed from $b_n$ in Figure \ref{fig:bnWignereven2} to confirm the exponential growth for $N_W = 1600$. For comparison, we also plot $K(t)$ computed from the linear fitting of $b_n$ where we exclude the fluctuation by hand (blue dashed curves). We also plot $K(t)$ for the power spectrum
\begin{align}\label{GaplessSpectrum}
f(\omega) & = \mathcal{N} e^{-\vert\omega\vert/\Omega},
\end{align}
which exponentially decays without gaps (green dotted curves). Their properties are 

\begin{itemize}
\item
In Figure \ref{fig:KCWignereven2(a)}, the three curves are very similar since the fluctuation of $b_n$ for $\Omega = 10$ in Figure \ref{fig:bnWignereven2} is very small.
\item
In Figure \ref{fig:KCWignereven2(b)}, the slopes of blue and green curves in the log plot at late times are similar. However, the values of $K(t)$ are very different. This is because the Lanczos behaves as $b_n\sim \pi \Omega n/2 +\gamma$ for these curves are similar but the values of $\gamma$ are different between them. When $K(t)\sim\frac{\eta}{4} e^{\pi\Omega t}$, $\eta$ is related to $\gamma$ as $\eta\sim 4\gamma/(\pi\Omega)+1$ \cite{Parker:2018yvk}. In Figure \ref{fig:KCWignereven2(b)}, $\frac{\eta}{4}$ for the blue curve is $\frac{\eta}{4}\sim2.0$, and  $\frac{\eta}{4}$ for the green curve is $\frac{\eta}{4}\sim0.25$.
\item
The slope of the red curve in \ref{fig:KCWignereven2(b)} at late times is smaller than the one of the blue and green curves. This is because of the fluctuation of $b_n$. The fluctuation in $b_n$ makes the slope smaller as we will see later in more detail. However, since the red solid curve in \ref{fig:KCWignereven2(b)} grows linearly in the log plots at late times, $K(t)$ for the spectra with infinitely many Wigner semicircles would grow exponentially. This leads to the conclusion that the fluctuations of $b_n$ make the exponential growth milder, but still, it maintains the exponential form as $K(t) \sim \exp\left(\alpha t \right)$ but with smaller $\alpha$ by fluctuations.  Such a mild exponential growth of $K(t)$ was observed in \cite{Avdoshkin:2022xuw,Camargo:2022rnt} for the spectrum that has a single mass gap, and our numerical computation suggests that a similar phenomenon happens even when the spectrum has an infinite number of gaps.
\item
If the number of Wigner semicircles, $N_W$ is finite, then $b_n$ saturates as in Figure \ref{fig:bnWignereven} at some $n$, and then the growth of $K(t)$ transitions to the linear growth at late times. Our numerical computations imply that, unless the power spectrum is discrete, the existence of gaps changes the coefficient of exponential growth of $K(t)$, but does maintain the exponential growth of the Krylov complexity. 
\end{itemize}

\subsubsection{Krylov complexity for Wigner semicircles with various $2\ell \le m$}

\begin{enumerate}
\item {\bf \underline{Small $\ell \to 0$}}\\
Let us consider the small $\ell$ and also the limit $\ell \to 0$. We consider the spectral density given by eq.~\eqref{SpectraSemicircles}, \eqref{SpectraSemicircles2} with $m=10$, $\Omega=1/12$, $N_W=25$ (51 semicircles), and $\ell =1/10$ and $\ell = 1/100$ respectively. 

The smaller $\ell$ is, the closer the spectrum is to a discrete spectrum. 
Figure \ref{fig:bnWignereven3} show the Lanczos coefficient $b_n$ for the above parameters. Comparing Figures \ref{fig:bnWignereven3(a)} and \ref{fig:bnWignereven3(b)}, the smaller $\ell$ is, the larger fluctuation of $b_n$. This implies that the fluctuation is large if the spectrum is closer to a discrete spectrum. Figure \ref{fig:KCWignereven3} shows the Krylov complexity $K(t)$ computed from $b_n$ for various small $\ell$. We plot two figures with different horizontal scales, where the maximum value of $t$ in the left figure is $t_{max}=20000$, and $t_{max}=200000$ in the right figure. One can see that $K(t)$ with the same value of $\ell t_{max}$ behaves similarly in these figures. This is because (\ref{C(t)finiteWigner}) depends on $\ell$ in the form of $\ell t$.
From Figures \ref{fig:bnWignereven3} and \ref{fig:KCWignereven3}, we can see that the larger the fluctuation of $b_n$, the smaller the slope of $\log K(t)$.
Thus, in the limit $\ell \to 0$, huge fluctuations kill the exponential growth of $K(t)$ and we have a transition to the oscillation behavior of $K(t)$ in that limit.

\begin{figure}[t]
\centering
     \begin{subfigure}[b]{0.4\textwidth}
         \centering
         \includegraphics[width=\textwidth]{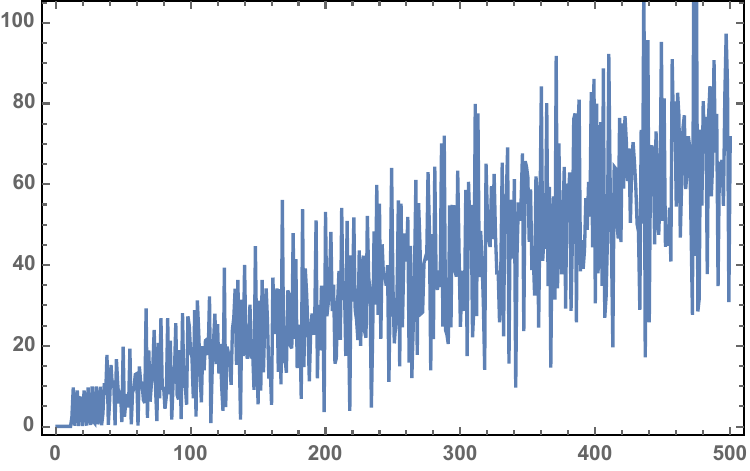}
       \put(5,0){$n$}
    \put(-175,115){$b_n$}
       \caption{$b_n$ for $\ell = 1/10$.}\label{fig:bnWignereven3(a)}
     \end{subfigure}
      \quad\quad\quad
     \begin{subfigure}[b]{0.4\textwidth}
         \centering
         \includegraphics[width=\textwidth]{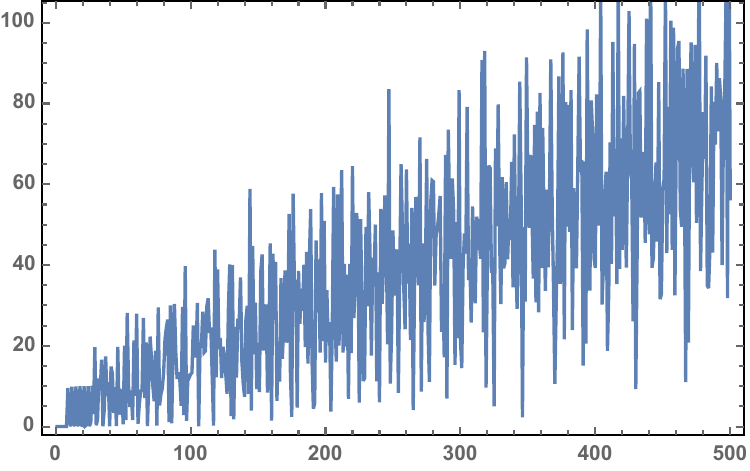}
          \put(5,0){$n$}
    \put(-175,115){$b_n$}
         \caption{$b_n$ for $\ell = 1/100$.}\label{fig:bnWignereven3(b)}
     \end{subfigure}
             \caption{Lanczos coefficients $b_n$ for the even spectra with small $\ell$.}
        \label{fig:bnWignereven3}
\end{figure}

\begin{figure}[t]
\centering
     \begin{subfigure}[b]{0.4\textwidth}
         \centering
         \includegraphics[width=\textwidth]{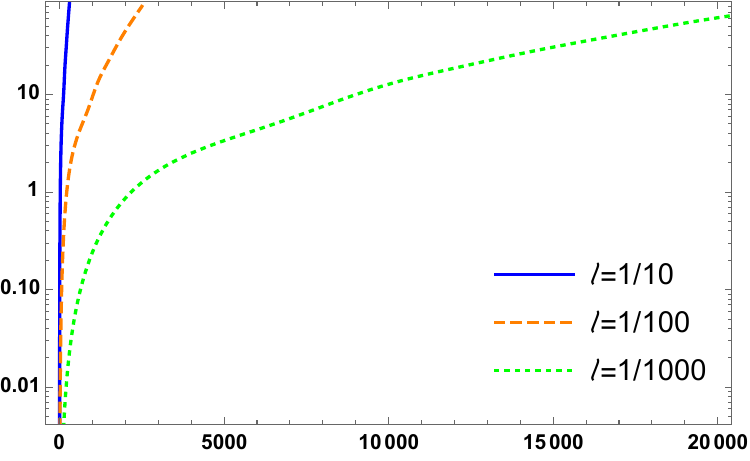}
       \put(5,0){$t$}
    \put(-175,115){$K(t)$}
     \caption{$K(t)$ from $t=0$ to $t = 20000$.}\label{}
     \end{subfigure}
      \quad\quad\quad
     \begin{subfigure}[b]{0.4\textwidth}
         \centering
         \includegraphics[width=\textwidth]{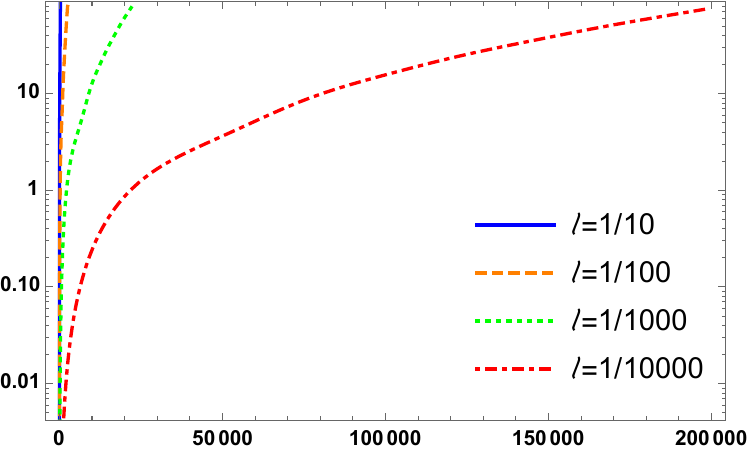}
       \put(5,0){$t$}
    \put(-175,115){$K(t)$}
     \caption{$K(t)$ from $t=0$ to $t = 200000$.}\label{}
     \end{subfigure}
                  \caption{Log plots of Krylov complexities $K(t)$ for various small $\ell$. We plot two figures with different horizontal scales of $t$.}
        \label{fig:KCWignereven3}
\end{figure}

\item {\bf \underline{Large $\ell \to m/2$}}\\ 
Let us study the fluctuation of $b_n$ when the Wigner semicircles in the spectrum touch with each other as in Figure \ref{fig:SpectrumWignerSemicirclesL5}, where $\ell \to m/2$. For visibility, we choose $\Omega=10$ in Figure \ref{fig:SpectrumWignerSemicirclesL5}. In Figure \ref{fig:bnWignerSemicirclesL5} with $\Omega=1/12$ for the strong exponential decay of $A_j$, we can see that the fluctuation of $b_n$ exists even where $\ell \to m/2$. This is because the spectrum is not smooth between the Wigner semicircles.

\begin{figure}[t]
\centering
     \begin{subfigure}[b]{0.4\textwidth}
         \centering
         \includegraphics[width=\textwidth]{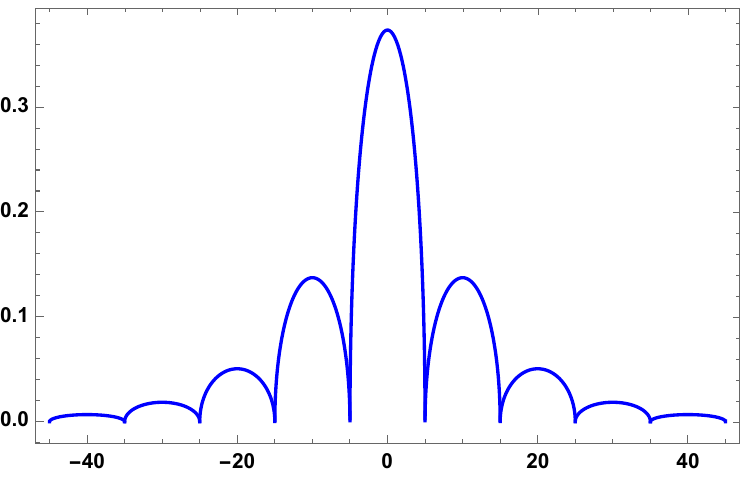}
     \put(5,5){$\omega$}
    \put(-175,115){$f(\omega)$}
       \caption{$f(\omega)$ for $\Omega = 10$.}\label{fig:SpectrumWignerSemicirclesL5}
     \end{subfigure}
      \quad\quad\quad
     \begin{subfigure}[b]{0.4\textwidth}
         \centering
         \includegraphics[width=\textwidth]{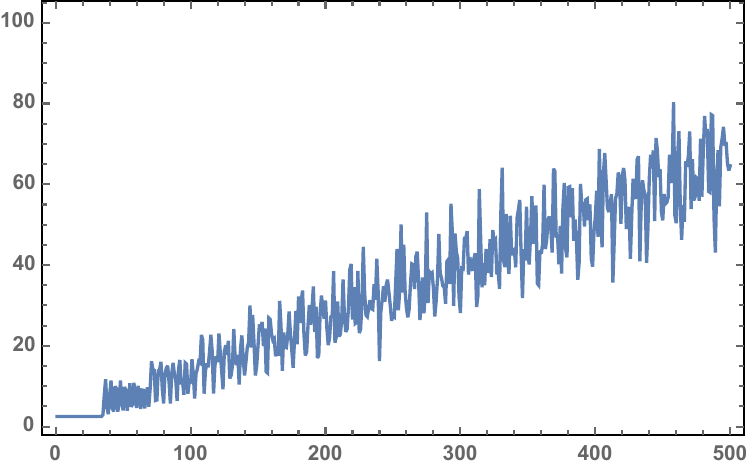}
          \put(5,0){$n$}
    \put(-175,115){$b_n$}
         \caption{$b_n$ for $\Omega=1/12$.}\label{fig:bnWignerSemicirclesL5}
     \end{subfigure}
             \caption{Spectrum $f(\omega)$ and Lanczos coefficient $b_n$ with $m=10$, $N_W=25$, $\ell=5$.}
        \label{fig:bnWignereven4}
\end{figure}

\end{enumerate}

\subsubsection{Nonsymmetric model $a_n \neq 0$}
\begin{figure}[h]
         \centering
         \includegraphics[width=0.5\textwidth]{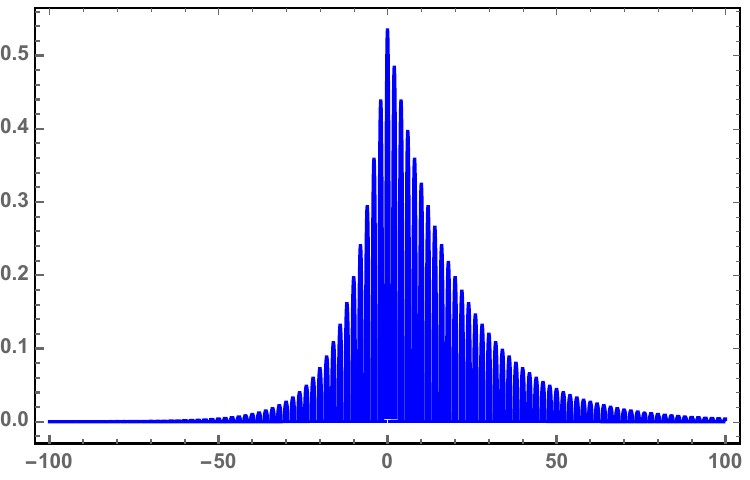}
       \put(5,5){$\omega$}
    \put(-225,143){$f(\omega)$}
       \caption{Non-symmetric spectrum $f(\omega)$ (\ref{SpectraSemicircles}) with $m=2, \ell =1/2, \Omega_+=20, \Omega_-=10, N_W=50$.}\label{fig:NonSymmetricSpectrumWignerSemicircles}
\end{figure}

At finite temperature $y\ne1$, our key equation eq.~(\ref{real}) is not symmetric under $\omega\to-\omega$. Due to this non-symmetric property of the equation, the decay rate of $F(\omega)$ is not symmetric under $\omega\to-\omega$. To see this non-symmetric effect, let us consider the spectrum (\ref{SpectraSemicircles}) with
\begin{align}\label{NonSymmetricSpectrum}
 \omega_j = m j  \,, \quad \quad A_j=
\begin{cases}
e^{-\vert\omega_j\vert/\Omega_+} & (j\ge0) \\
e^{-\vert\omega_j\vert/\Omega_-} & (j<0)
\end{cases}
.
\end{align}
For example, Figure \ref{fig:NonSymmetricSpectrumWignerSemicircles} is a plot of the non-symmetric spectrum with $m=2, \ell =1/2, \Omega_+=20, \Omega_-=10, N_W=50$.

\begin{figure}[t]
\centering
     \begin{subfigure}[b]{0.4\textwidth}
         \centering
         \includegraphics[width=\textwidth]{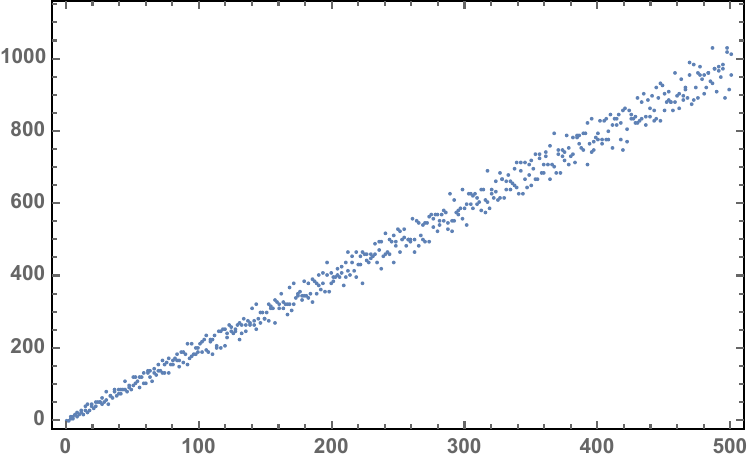}
       \put(5,0){$n$}
    \put(-170,113){$a_n$}
     \end{subfigure}
      \quad\quad\quad
     \begin{subfigure}[b]{0.4\textwidth}
         \centering
         \includegraphics[width=\textwidth]{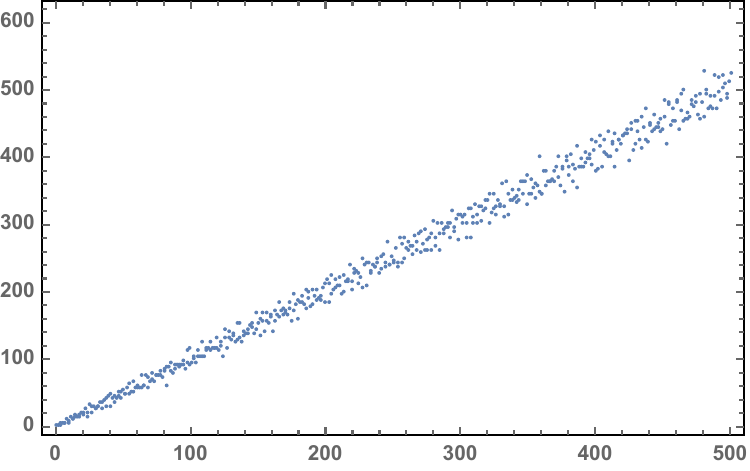}
          \put(5,0){$n$}
    \put(-170,113){$b_n$}
     \end{subfigure}
             \caption{Lanczos coefficients $a_n$ and $b_n$ of the non-symmetric spectrum with $m=10, \ell =2, \Omega_+=1, \Omega_-=1/12, N_W=200$.}
        \label{fig:anbnNonSymmetricWignerSemicircles}
\end{figure}

\begin{figure}[t]
\centering
     \begin{subfigure}[b]{0.4\textwidth}
         \centering
         \includegraphics[width=\textwidth]{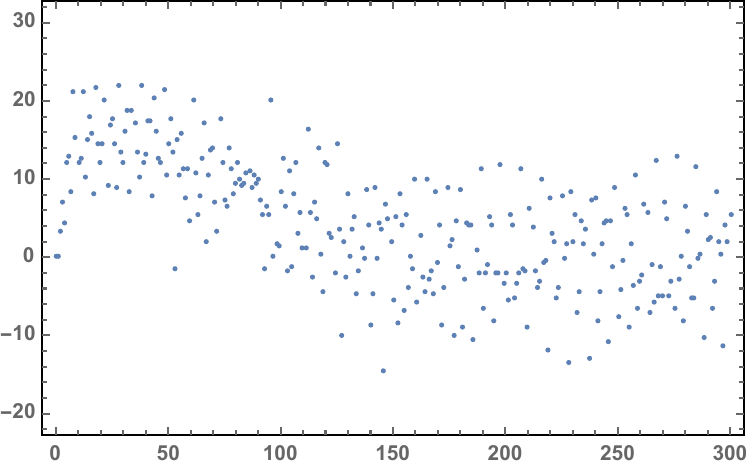}
       \put(5,0){$n$}
    \put(-170,113){$a_n$}
     \end{subfigure}
      \quad\quad\quad
     \begin{subfigure}[b]{0.4\textwidth}
         \centering
         \includegraphics[width=\textwidth]{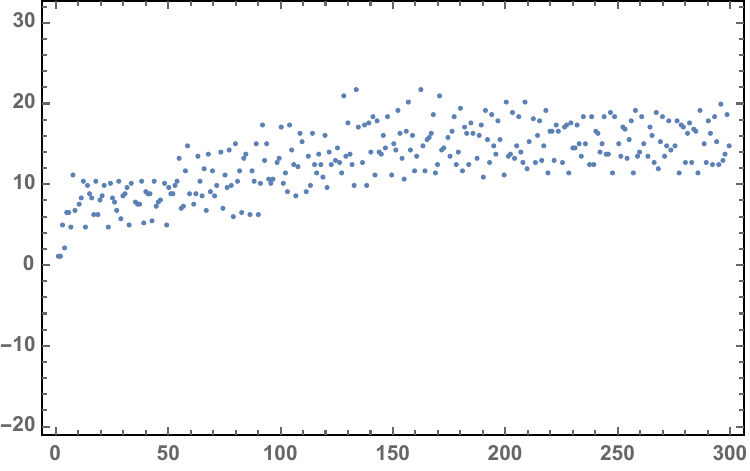}
          \put(5,0){$n$}
    \put(-170,113){$b_n$}
     \end{subfigure}
             \caption{Lanczos coefficients $a_n$ and $b_n$ of the non-symmetric spectrum with $m=10, \ell =2, \Omega_+=1, \Omega_-=1/12, N_W=3$.}
        \label{fig:anbnNonSymmetricWignerSemicirclesNw3}
\end{figure}

\begin{figure}[t]
\centering
     \begin{subfigure}[b]{0.5\textwidth}
         \centering
         \includegraphics[width=\textwidth]{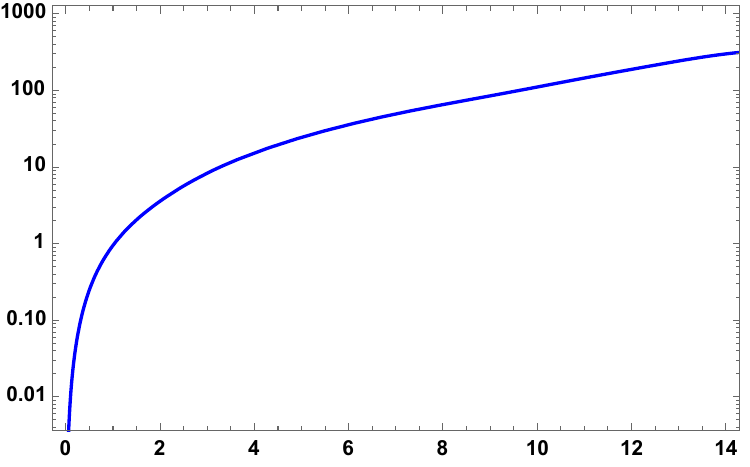}
       \put(5,0){$t$}
    \put(-225,145){$K(t)$}
     \end{subfigure}
                  \caption{Log plot of Krylov complexity $K(t)$ for the non-symmetric spectrum with $m=10, \ell =2, \Omega_+=1, \Omega_-=1/12, N_W=200$.}
        \label{fig:KCNonSymmetricWigner}
\end{figure}

We numerically compute the Lanczos coefficients $a_n$ and $b_n$ of this non-symmetric spectrum with $m=10$, $\ell =2$, $\Omega_+=1$, $\Omega_-=1/12$, $N_W=200$ as shown in Figure \ref{fig:anbnNonSymmetricWignerSemicircles}. In this numerical computation, the Lanczos coefficients increase linearly with fluctuation. If $N_W$ is small, as shown in Figure \ref{fig:anbnNonSymmetricWignerSemicirclesNw3}, $a_n$ fluctuates around $a_n=0$ at large $n$, and $b_n$ saturates to a nonzero value with fluctuation. Figure \ref{fig:KCNonSymmetricWigner} is a log plot of Krylov complexity for the non-symmetric spectrum with $N_W=200$ computed from the Lanczos coefficients in Figure \ref{fig:anbnNonSymmetricWignerSemicircles}. It shows the exponential growth of $K(t)$ at late times. Therefore, our numerical computation of the non-symmetric model indicates that the non-symmetric decay rate of $F(\omega)$ does not change the exponential growth of Krylov complexity.

In summary, our study of the low-temperature gapped spectral density model shows that 
\begin{enumerate}
\item At $\ell = 0$ corresponds to $T=0$, the Krylov complexity oscillates and does not grow. 
\item However once $\ell > 0$ corresponds to $T > 0$, the Lanczos coefficients grow linearly in $n$ with fluctuations. Fluctuations of $b_n$ are huge near $\ell \to 0$ but as we increase $\ell$, the fluctuations become smaller. 
\item Fluctuations of $b_n$ makes the slope of $\log K(t)$ smaller. However, as long as $\ell \neq 0$, the Lancoz coefficients $b_n$ grow linearly with fluctuations and the Krylov complexity grow exponentially as $K(t) \sim \exp\left(\alpha t \right)$. In the limit $\ell \to 0$, $\alpha \to 0$ and as we increases $\ell$, $\alpha$ also increases. 
\item It is crucial for the $K(t)$ grows exponentially in time, or equivalently for $b_n$ grows linearly in $n$, there are infinite cuts, {\it i.e.,} $N_W \to \infty$. For finite $N_W$, the exponential growth of $K(t)$ saturates at some time. $N_W \to \infty$ is the nature of the key equation eq.~\eqref{real} obtained from the Schwinger-Dyson equation of the IP model in the large $N$ limit. 
\end{enumerate}

\subsection{High-temperature $T > T_c$ model for the gapless spectrum with peaks}
The spectrum becomes gapless at high temperatures due to the merging of branch cuts. Near the temperature at which the spectrum becomes gapless, the gapless spectrum has multiple peaks as shown in Figure \ref{fig:F(w)} for $m=0.2, \nu_T=1, y=0.25$, which is the remnant of the multiple cuts. 
Since we have already seen that the Krylov complexity grows exponentially even in the presence of gaps, it is quite reasonable to guess that for all nonzero temperature $T > 0$, the Krylov complexity grows exponentially in time.

\begin{figure}
         \centering
         \includegraphics[width=0.5\textwidth]{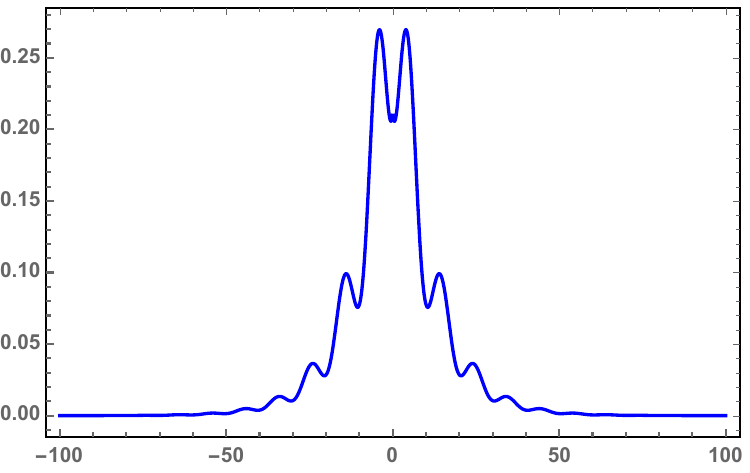}
       \put(5,5){$\omega$}
    \put(-225,140){$f(\omega)$}
       \caption{High temperature model $f(\omega)$ (\ref{SpectrumPeaks}) with $m=10, \Omega=10$.}\label{fig:SymmetricSpectrumPeaks}
\end{figure}

\begin{figure}[t]
\centering
     \begin{subfigure}[b]{0.5\textwidth}
         \centering
         \includegraphics[width=\textwidth]{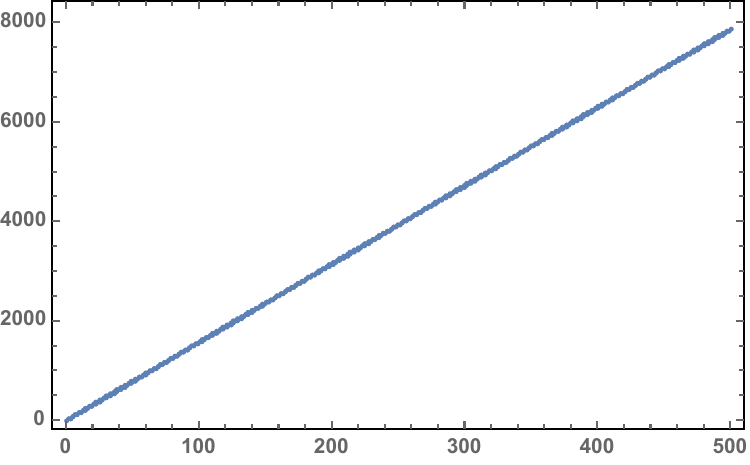}
          \put(5,0){$n$}
    \put(-215,140){$b_n$}
     \end{subfigure}
             \caption{Lanczos coefficient $b_n$ of the high-temperature model $f(\omega)$ (\ref{SpectrumPeaks}) with $m=10, \Omega=10$. This plot is almost identical to the one in Figure \ref{fig:bnWignereven2},}
        \label{fig:bnSymmetricSpectrumPeaks}
\end{figure}

To study the effect of these peaks on Krylov complexity, we introduce the following model
\begin{align}\label{SpectrumPeaks}
f(\omega)=\mathcal{N}(1+\sin^2(\pi \omega/m))e^{-\vert \omega\vert/\Omega}.
\end{align}
Figure \ref{fig:SymmetricSpectrumPeaks} is a plot of this toy model with $m=10, \Omega=10$, which shows peaks due to $\sin^2(\pi \omega/m)$. This phase is chosen such that it is periodicic $\omega \to \omega \pm m$. We numerically calculate the Lanczos coefficient $b_n$ of this model as shown in Figure  \ref{fig:bnSymmetricSpectrumPeaks}. The linear increase behavior of $b_n$ in Figure \ref{fig:bnSymmetricSpectrumPeaks} is almost identical to the one with a large number of Wigner semicircles in Figure \ref{fig:bnWignereven2}, where the exponential decay rate $\Omega$ in these figures is the same value. Therefore, from our numerical computation, we conclude the multiple-peaks in the gapless spectrum associated with the multiple cuts at small temperatures are not so relevant to the linear increase of the Lanczos coefficient. Thus, as is expected, the Krylov complexity grows exponentially at high temperatures where the spectrum becomes gapless.

Finally, we comment on the difference of exponential behaviors of $K(t)$ between $e^t$ and $e^{\sqrt{t}}$ due to the $\log n$ correction of $b_n$: Here we neglected log corrections in the Lanczos coefficient, which makes the IP model complexity $e^{\sqrt{t}}$ instead of $e^t$. 

Finally, we comment on the difference between $\exp(\mathcal{O}(t))$ and $\exp(\mathcal{O}(\sqrt{t}))$ in $K(t)$. In this section, we have focused our analysis on the difference between an exponential growth in $K(t)$ and growth in the power law. Therefore, we have not paid much attention to the difference between $\exp(\mathcal{O}(t))$ and $\exp(\mathcal{O}(\sqrt{t}))$. This difference is due to the log correction in the Lanczos coefficient $b_n$, and in the actual IP model, $K(t)$ is always $\exp(\mathcal{O}(\sqrt{t}))$ due to this correction.


\section{Numerical analysis of the IP model at finite temperature}
\label{mainnumerics}

In this section, we numerically compute the Lanczos coefficients by using $F(\omega)$ of the IP model at finite temperature as shown in Figure \ref{fig:F(w)}. Since $F(\omega)$ of the IP model decays exponentially at large $\vert \omega\vert$, the numerical calculation of $F(\omega)$ at large $\vert \omega\vert$ with high accuracy is difficult. For this reason, we introduce a cutoff scale $\omega_c$ and numerically construct a continuous function $F(\omega)$ by solving (\ref{real}) with a condition $F(\omega)=0$ in $\vert\omega\vert>\omega_c$. To solve the difference equation, we use a boundary condition $\tilde{G}(T,\omega)=\tilde{G}_0$ (\ref{freepropagator}) around $\vert\omega\vert=\omega_c$ since the IP model becomes a free theory at UV.

Due to the above difficulty regarding the accuracy of numerical calculations, we can only do numerical calculations with small cutoff $\omega_c$ for $N_W\lesssim10$. The results of such numerical computations with the small cutoff are expected to behave like Figure \ref{fig:anbnNonSymmetricWignerSemicirclesNw3} for small $N_W$ rather than Figure \ref{fig:anbnNonSymmetricWignerSemicircles} for large $N_W$.

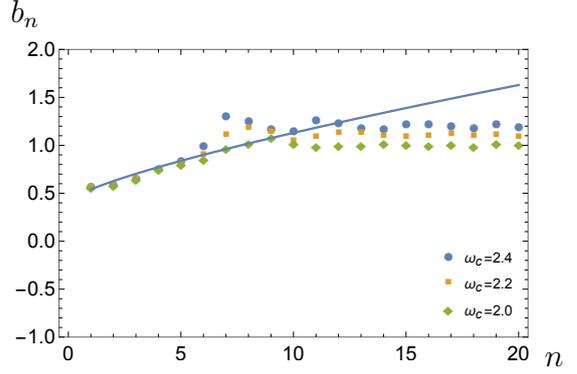
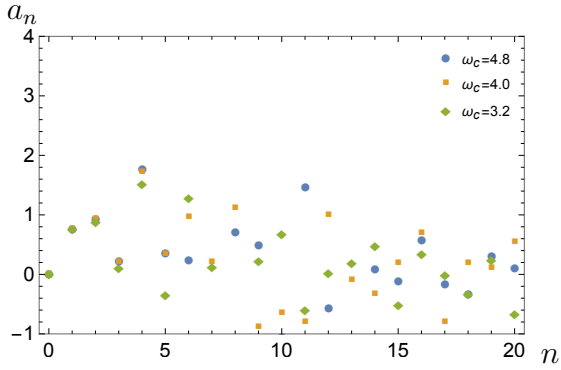
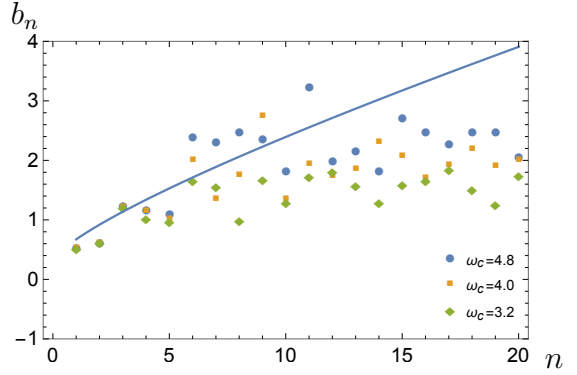
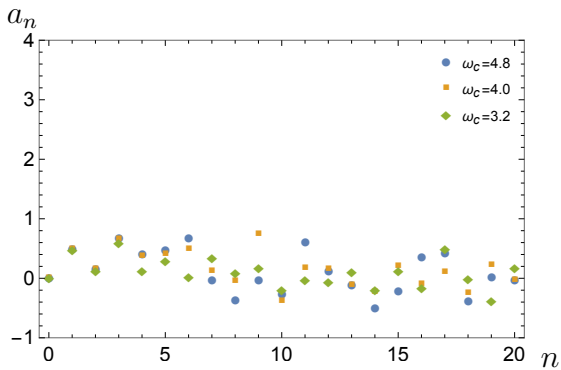
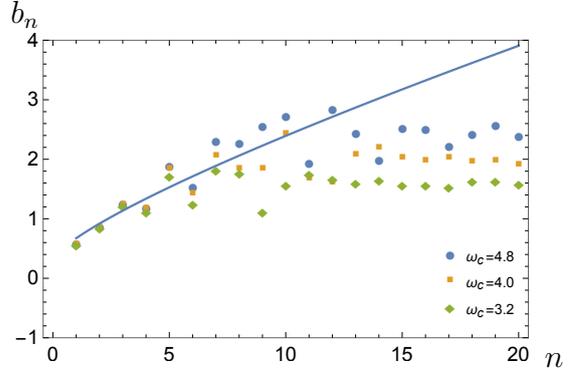
\begin{figure}
     \begin{subfigure}[b]{0.45\textwidth}
         \centering
         \includegraphics[width=\textwidth]{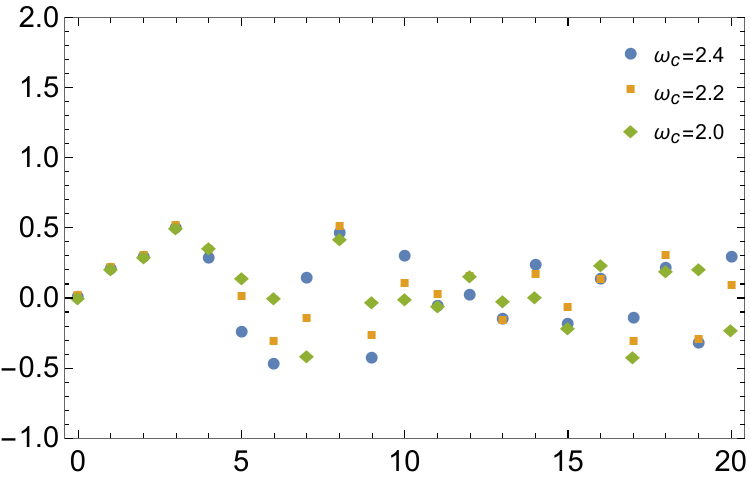}
       \put(5,0){$n$}
    \put(-195,130){$a_n$}
       \caption{$a_n$ for $m=0.2, \;\nu_T=1, \;y=0.04$.}\label{fig:LanczosIPmodel(a)}
     \end{subfigure}
      \hfill
     \begin{subfigure}[b]{0.45\textwidth}
         \centering
         \includegraphics[width=\textwidth]{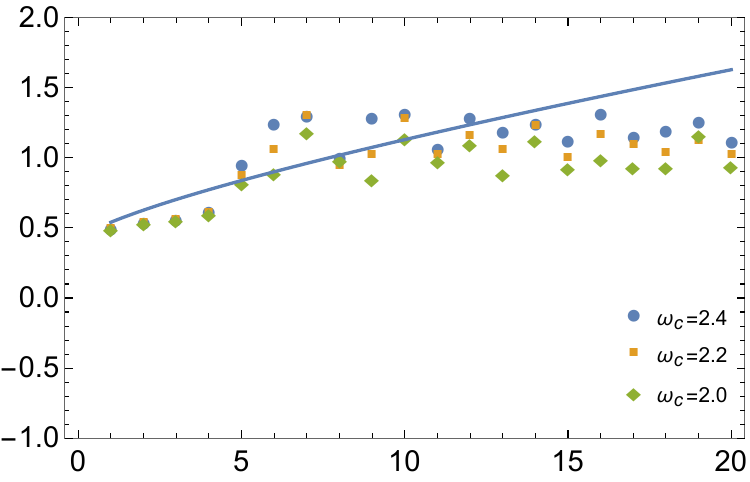}
          \put(5,0){$n$}
    \put(-195,130){$b_n$}
         \caption{$b_n$ for $m=0.2, \;\nu_T=1, \;y=0.04$.}\label{fig:LanczosIPmodel(b)}
     \end{subfigure}
          \begin{subfigure}[b]{0.45\textwidth}
         \centering
         \includegraphics[width=\textwidth]{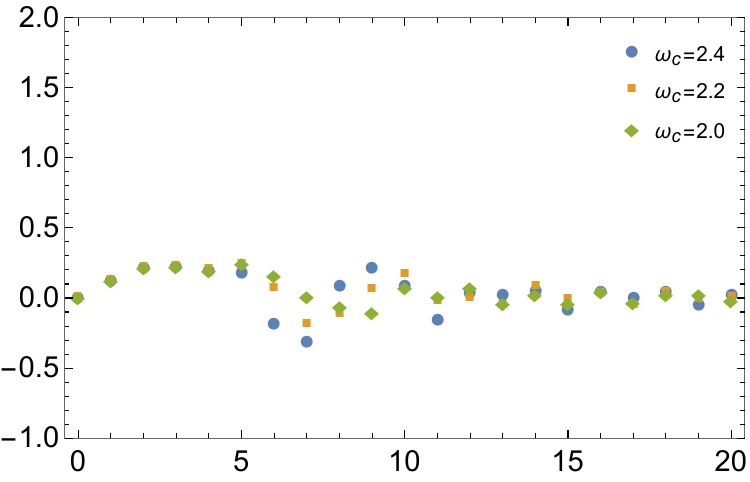}
       \put(5,0){$n$}
    \put(-195,130){$a_n$}
       \caption{$a_n$ for $m=0.2, \;\nu_T=1, \;y=0.25$.}\label{fig:LanczosIPmodel(c)}
     \end{subfigure}
      \hfill
     \begin{subfigure}[b]{0.45\textwidth}
         \centering
         \includegraphics[width=\textwidth]{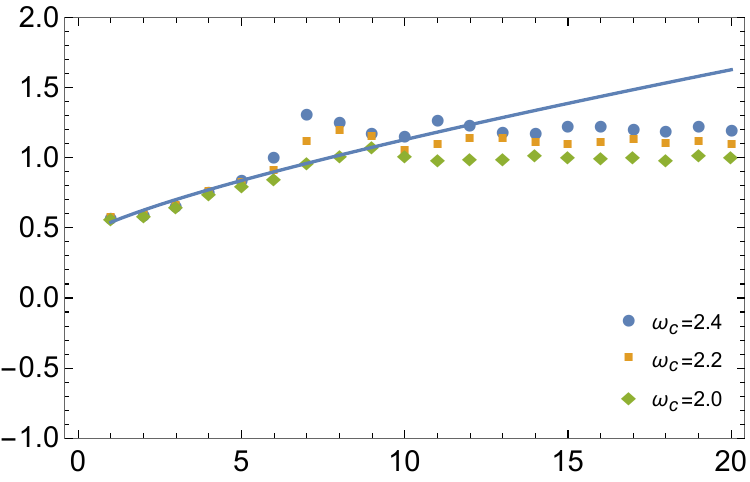}
          \put(5,0){$n$}
    \put(-195,130){$b_n$}
         \caption{$b_n$ for $m=0.2, \;\nu_T=1, \;y=0.25$.}\label{fig:LanczosIPmodel(d)}
     \end{subfigure}
          \begin{subfigure}[b]{0.45\textwidth}
         \centering
         \includegraphics[width=\textwidth]{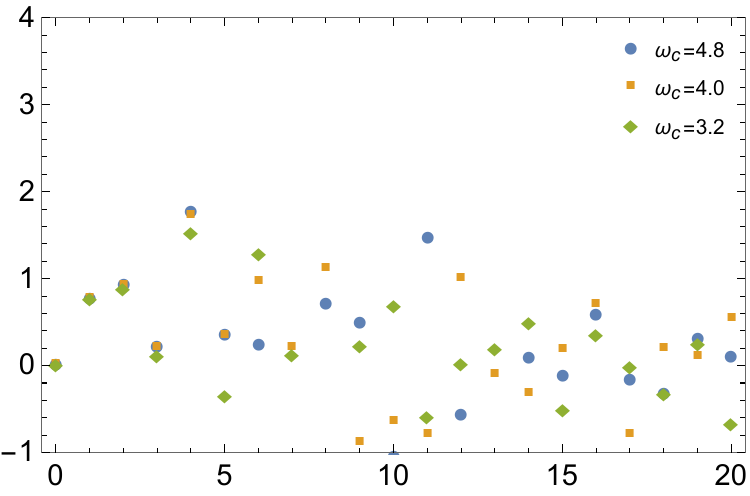}
       \put(5,0){$n$}
    \put(-195,130){$a_n$}
       \caption{$a_n$ for $m=0.8, \;\nu_T=1, \;y=0.04$.}\label{fig:LanczosIPmodel(e)}
     \end{subfigure}
      \hfill
     \begin{subfigure}[b]{0.45\textwidth}
         \centering
         \includegraphics[width=\textwidth]{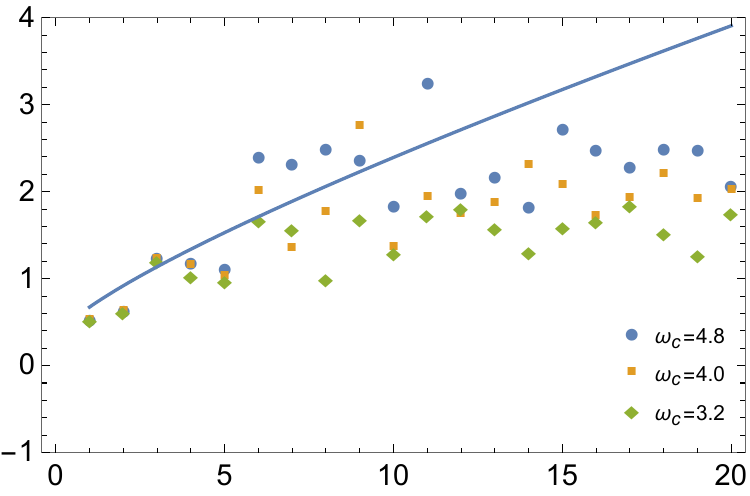}
          \put(5,0){$n$}
    \put(-195,130){$b_n$}
         \caption{$b_n$ for $m=0.8, \;\nu_T=1, \;y=0.04$.}\label{fig:LanczosIPmodel(f)}
     \end{subfigure}
          \begin{subfigure}[b]{0.45\textwidth}
         \centering
         \includegraphics[width=\textwidth]{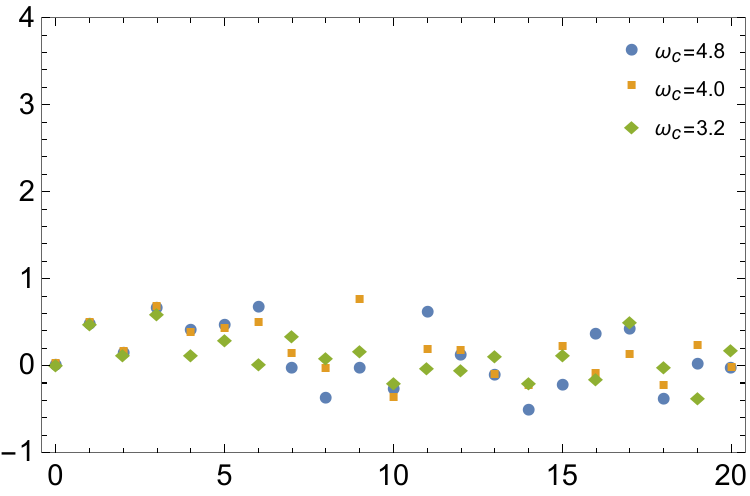}
       \put(5,0){$n$}
    \put(-195,130){$a_n$}
       \caption{$a_n$ for $m=0.8, \;\nu_T=1, \;y=0.25$.}\label{fig:LanczosIPmodel(g)}
     \end{subfigure}
      \hfill
     \begin{subfigure}[b]{0.45\textwidth}
         \centering
         \includegraphics[width=\textwidth]{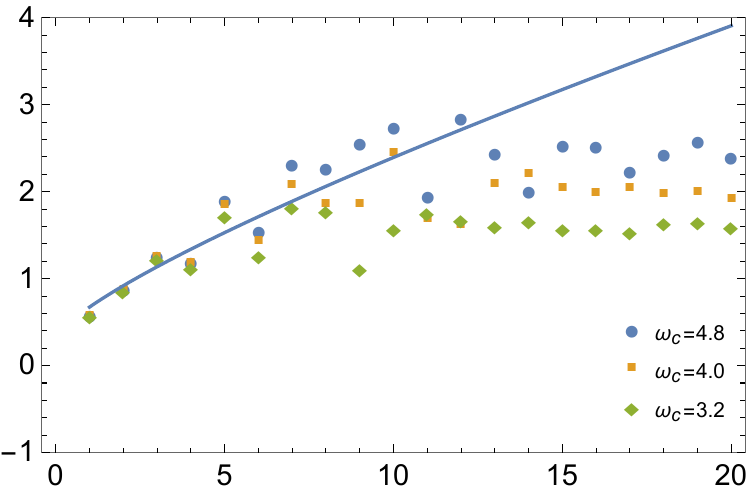}
          \put(5,0){$n$}
    \put(-195,130){$b_n$}
         \caption{$b_n$ for $m=0.8, \;\nu_T=1, \;y=0.25$.}\label{fig:LanczosIPmodel(h)}
     \end{subfigure}
             \caption{Lanczos coefficients of the IP model at finite temperature, where $y=e^{-m/T}$ and $\omega_c$ is a cutoff such that $F(\omega)=0$ in $\vert\omega\vert>\omega_c$. We also plot curves $b_n=b_0+ \frac{m\pi n}{4W(2m\pi n/\nu_T)}$, where $b_0=0.3$ for $m=0.2$ and $b_0=0.2$ for $m=0.8$.
             }
        \label{fig:LanczosIPmodel}
\end{figure}

Figure \ref{fig:LanczosIPmodel} shows numerical plots of the Lanczos coefficients of the IP model at finite temperature. As $n$ increases, the Lanczos coefficient $b_n$ begins to saturate like $b_n\sim\omega_c/2$, which is similar to the behavior of $b_n$ in the IP model at infinite temperature with finite $\omega_c$ \cite{Iizuka:2023pov}. When the cutoff is infinite $\omega_c\to\infty$, $b_n$ would  grow without the saturation. For comparison with the increasing behavior (\ref{abbnIP}) at infinite temperature $T$, we also plot $b_n=b_0+ \frac{m\pi n}{4W(2m\pi n/\nu_T)}$, where $b_0$ is a constant. The Lanczos coefficient $a_n$ at large $n$ with finite $\omega_c$ fluctuates around $a_n=0$. This may be related to our symmetric setting of the cutoff $\omega_c$ since $a_n$ is zero for the symmetric spectrum. These properties are qualitatively consistent with the behaviors in Figure \ref{fig:anbnNonSymmetricWignerSemicirclesNw3}.

Another important observation from these numerical plots is that the fluctuation of Lanczos coefficients becomes larger as temperature $T$ decreases, where the low temperature leads to small $y$. This behavior is consistent with the behavior of our toy model in Section \ref{mainanalysis} as follows. The spectrum of the IP model at nonzero low temperature can be approximated by the infinite sum of Wigner semicircles whose length $\ell$ becomes smaller as the temperature decreases. As we have seen in Figure \ref{fig:bnWignereven3}, the fluctuation of $b_n$ becomes larger as $\ell$ decreases, which is similar to the behavior of $b_n$ in Figure \ref{fig:LanczosIPmodel}. Thus, our toy model in Section \ref{mainanalysis} nicely captures the IP model's characteristic that the fluctuation of Lanczos coefficients becomes larger as $T$ decreases.

For the precise study of the large-$n$ behavior of Lanczos coefficients and the late-time behavior of Krylov complexity, it is important to perform accurate numerical computations with large cutoff $\omega_c$, and we leave it as a future work. It is also interesting to carefully examine the effect of nonzero $a_n$ on the Krylov complexity.

\newpage
\clearpage

\newpage
\section{Krylov complexity in the IOP matrix model}
\label{IOPKrylov}

\subsection{Planar limit}
To understand the Krylov complexity for other models, in this section, we study it in the IOP model \cite{Iizuka:2008eb}. The IOP model Hamiltonian is 
\begin{eqnarray}
H = \frac{1}{2} {\rm Tr}(\Pi^2) + \frac{m^2}{2} {\rm Tr}(X^2) + Ma_i^\dagger a_i 
+ h a_i^\dagger a_l  A^\dagger_{ij} A_{jl}  \ , \label{IOPH}
\end{eqnarray}
Note that contrary to the IP model, the interaction term will not change the adjoint $A^\dagger$ excitations.    

In the large $N$ planar limit, the Schwinger-Dyson equation for the fundamental can be solved and we obtain  \cite{Iizuka:2008eb}
\begin{equation}
\tilde G(T,\omega)= \frac{i(1-y)}{2\omega\lambda}\left( \lambda + \omega - \sqrt{(\omega-\omega_+)
(\omega-\omega_-)} \right)\ ,\quad
\omega_{\pm} = \lambda \frac{ 1+ y \pm 2 \sqrt{y} }{1-y}\ . \label{primed-G-SD}
\end{equation}
where $y= e^{-m/T}$ and $\lambda := h N$ is the 't Hooft coupling. 
Then, the spectral density $ \pho(\omega)$  is given by
\begin{align}\label{spectrumIOP}
\pho(\omega)&= \frac{1}{\pi}{\rm Re}\,\tilde G(T,\omega)  = \frac{(1-y)}{2} \delta(\omega)+\frac{1-y}{2 \pi \omega\lambda} \text{Re}\left[\sqrt{(\omega_+-\omega)(\omega-\omega_-)}\right] \,. 
\end{align}
Thus, the two-point function $C(t)$ with respect to time $t$ is defined by
\begin{align}
C(t)  :=\int^\infty_{-\infty} \frac{d \omega}{2\pi}e^{-i\omega t}f(\omega) \,, \quad f(\omega) = 2 \pi \pho(\omega)\,,
\end{align}
by eq.~\eqref{Ffrela}.
By using (\ref{Ci(t)}), we evaluate $\frac{d C(t)}{dt}$ as
\begin{align}
\frac{d C(t)}{dt}&=-i \int^\infty_{-\infty} \frac{d \omega}{2\pi}e^{-i\omega t}\omega f(\omega)=-i\frac{1-y}{\lambda}\int^{\omega_0+\ell_0}_{\omega_0-\ell_0} \frac{d \omega}{2\pi}e^{-i\omega t}\sqrt{\ell_0^2-(\omega-\omega_0)^2}\notag\\
&=-i\frac{(1-y)\ell_0}{2\lambda t}J_1(\ell_0 t)e^{-i \omega_0 t} \label{dtC(t)} \,,
\end{align}
where we set 
\begin{align}
\ell_0:&=\frac{\omega_+-\omega_-}{2},\;\;\; \omega_0:=\frac{\omega_++\omega_-}{2} \,.
\end{align}
$C(t)$ can be obtained by an integral of (\ref{dtC(t)}). Since the asymptotic behavior of (\ref{dtC(t)}) at late times is $t^{-3/2}$ with oscillation, its integral $C(t)$ at late times also has a power-law decay.  

\begin{figure}
         \centering
         \includegraphics[width=0.6\textwidth]{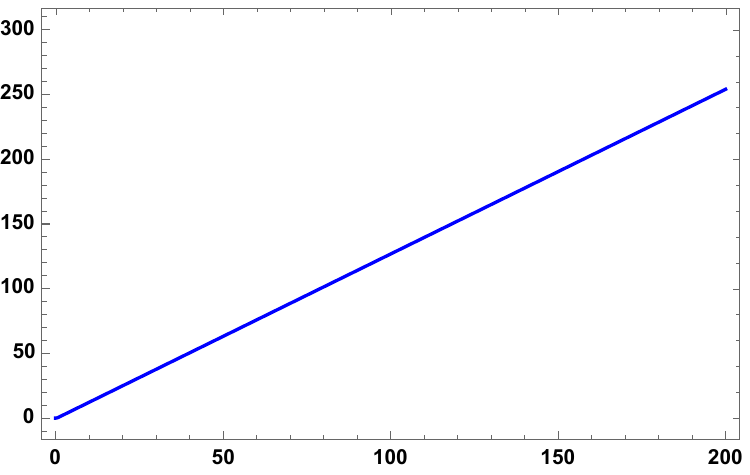}
       \put(5,5){$t$}
    \put(-265,165){$K(t)$}
       \caption{Planar limit Krylov complexity of the IOP model at infinite temperature with $\lambda'=1$.}\label{fig:KCIOPInfiniteT}
\end{figure}

We are interested in whether the IOP model at high temperature is chaotic and thus consider a limit $y\to1$ with fixed $\lambda':=\lambda/(1-y)$. In this limit, a pole disappears and 
$\omega_+ \to 4 \lambda'$  and 
$\omega_- \to 0$ 
and with these, $f(\omega)$ becomes 
\begin{align}\label{spectrumIOPInfiniteT}
f(\omega)=\text{Re}\left[\frac{1}{\omega\lambda'}\sqrt{(4\lambda'-\omega)\omega}\right] \,.
\end{align}
Its moment $M_n$ is obtained as 
\begin{align}
M_n:=\int^\infty_{-\infty} \frac{d \omega}{2\pi}\omega^nf(\omega)=\frac{2^{2 n}  \Gamma \left(n+\frac{1}{2}\right)}{\sqrt{\pi } \Gamma (n+2)} \lambda'^{n}.
\end{align}
The resultant Lanczos coefficients yielding this moment $M_n$ are obtained as 
\begin{align}\label{LanczosIOPInfiniteT}
a_0=\lambda', \;\;\; a_{n > 0 }=2\lambda' \;, \;\;\; b_n=\lambda'\,.
\end{align}
Since these Lanczos coefficients do not grow with $n$, the Krylov complexity does not grow exponentially in time in the IOP model.  
Figure \ref{fig:KCIOPInfiniteT} is a linear plot of Krylov complexity $K(t)$ computed from the Lanczos coefficients eq.~(\ref{LanczosIOPInfiniteT}) with $\lambda'=1$, which shows the linear growth of $K(t)$ in the IOP model at infinite temperature.

This is consistent with the results of the out-of-time-ordered correlator (OTOC) obtained \cite{Michel:2016kwn}, where OTOC does not show exponential growth. Although the direct relationship between the Krylov complexity and OTOCs are not yet fully understood, in the IOP model, neither Krylov complexity nor OTOC shows exponential growth in time.

\subsection{Non-planar corrections}
In \cite{Iizuka:2008eb}, the leading $1/N^2$ correction to the Green function was calculated. 
By expanding the correlator in $1/N^2$,
\begin{eqnarray}\label{GIOPN2}
\tilde{G}(T, \omega) = \tilde{G}^{(0)} (T, \omega) + { 1 \over N^2} \tilde{G}^{(1)} (T, \omega)+ {\cal{O}}\left({1 \over N^4}\right) \,, 
\end{eqnarray}
the leading term  $\tilde{G}^{(0)} (T, \omega)$ is given by eq.~\eqref{primed-G-SD} and 
the resultant subleading Green function $\tilde{G}^{(1)} (T, \omega)$ is given as 
\begin{equation} 
\label{finalsol}
\tilde{G}^{(1)}(T,\omega) =
\frac {i y^2 x_0^3  (1 -  x_0)^4 (1 - x_0 [1-y ] ) }
 {  (1 - 2 x_0 + x_0^2 [1-y])^4   (\omega [1-x_0]^2 - \lambda' y)} \,,  \quad x_0 := - i \lambda' \tilde{G}^{(0)} (T, \omega) \,.
 \end{equation}
One can immediately see that if both $\omega$ and $x_0$ are real values, then $\tilde{G}^{(1)}(T,\omega) $ is purely imaginary. Since for Re$\,\tilde{G}^{(0)}(T, \omega) = 0$, $x_0$ is the real value, this implies that  Re$\,\tilde{G}^{(1)}(T, \omega) \neq 0$ if and only if Re$\,\tilde{G}^{(0)}(T, \omega) \neq 0$, the branch cuts come only from the leading $\tilde{G}^{(0)}(T, \omega)$. 

\begin{figure}[t]
\centering
     \begin{subfigure}[b]{0.4\textwidth}
         \centering
         \includegraphics[width=\textwidth]{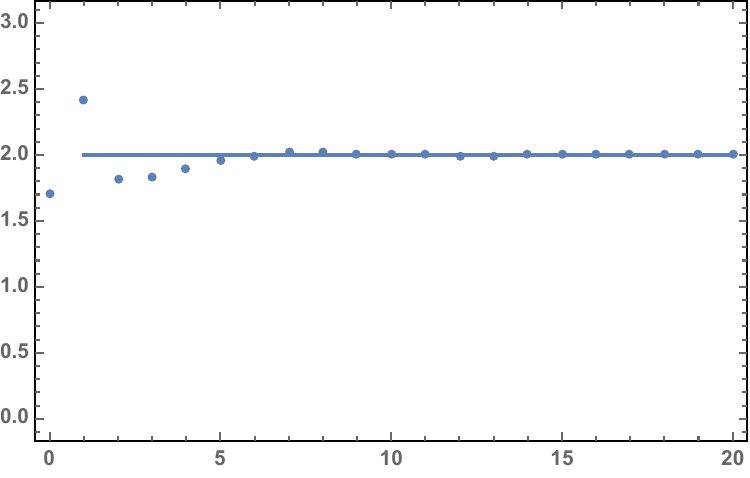}
       \put(5,0){$n$}
    \put(-175,115){$a_n$}
     \end{subfigure}
      \quad\quad\quad
     \begin{subfigure}[b]{0.4\textwidth}
         \centering
         \includegraphics[width=\textwidth]{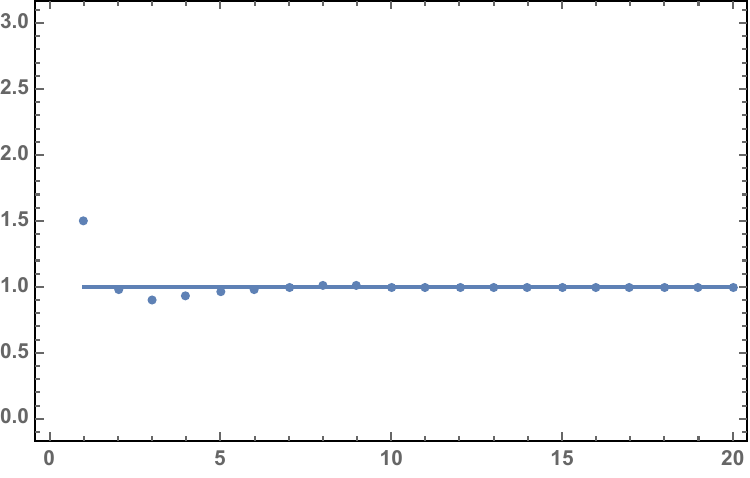}
          \put(5,0){$n$}
    \put(-175,115){$b_n$}
     \end{subfigure}
             \caption{Lanczos coefficients $a_n$ and $b_n$ of the IOP model up to the leading $1/N^2$ correction (\ref{GIOPN2}) with $y=1$, $\lambda'=1$, $N^2=100$, $\varepsilon=1/100$. Solid lines represent $a_{n>0} = 2 \lambda'=2$ and $b_n = \lambda'=1$ for comparison with the planar limit (\ref{LanczosIOPInfiniteT}).}
        \label{fig:anbnIOPN2}
\end{figure}

However, there is one difficulty to compute the Lanczos coefficients due to the singular behavior of $\tilde{G}^{(1)}(T,\omega)$, since $\tilde{G}^{(1)}(T,\omega)$ is singular at $\omega=\omega_{\pm}$ \cite{Iizuka:2008eb}. Thus, the following integral for $M_n$
\begin{align}
M_n=\int^{\omega_+}_{\omega_-} \frac{d \omega}{2\pi}\omega^nf(\omega)=\int^{\omega_+}_{\omega_-} \frac{d \omega}{\pi}\omega^n{\rm Re}\,\tilde G(T,\omega),
\end{align}
 does not converge. To regularize the integral, we introduce a cutoff $\varepsilon$ as
\begin{align}
M_n=\int^{\omega_+-\varepsilon}_{\omega_-+\varepsilon} \frac{d \omega}{\pi}\omega^n{\rm Re}\,\tilde G(T,\omega).
\end{align} 
By using this regularized integral, we compute the Lanczos coefficients with $y=1, \lambda'=1, N^2=100, \varepsilon=1/100$ as shown in Figure \ref{fig:anbnIOPN2}. One can see that the corrections of $a_n$ and $b_n$ from the planar limit are small at large $n$.

We have already seen that the spectral density is given by Re$\,\tilde{G}^{(0)}(T, \omega) $ has both upper and lower bound in $\omega$. Even after taking into account the $1/N^2$ corrections, the region where Re$\,\tilde{G}(T, \omega)$ is nonzero does not change. Thus the Lanczos coefficients cannot grow linearly in $n$ as seen in Figure \ref{fig:anbnIOPN2} if the integral of the spectrum is regularized. Thus, we conclude that the Krylov complexity of the IOP model does not grow exponentially in time even after taking into account the non-planar corrections as well as in the planar limit.


\section{Conclusions and discussions}
\label{conclusion}
\begin{figure}
         \centering
         \includegraphics[width=1\textwidth]{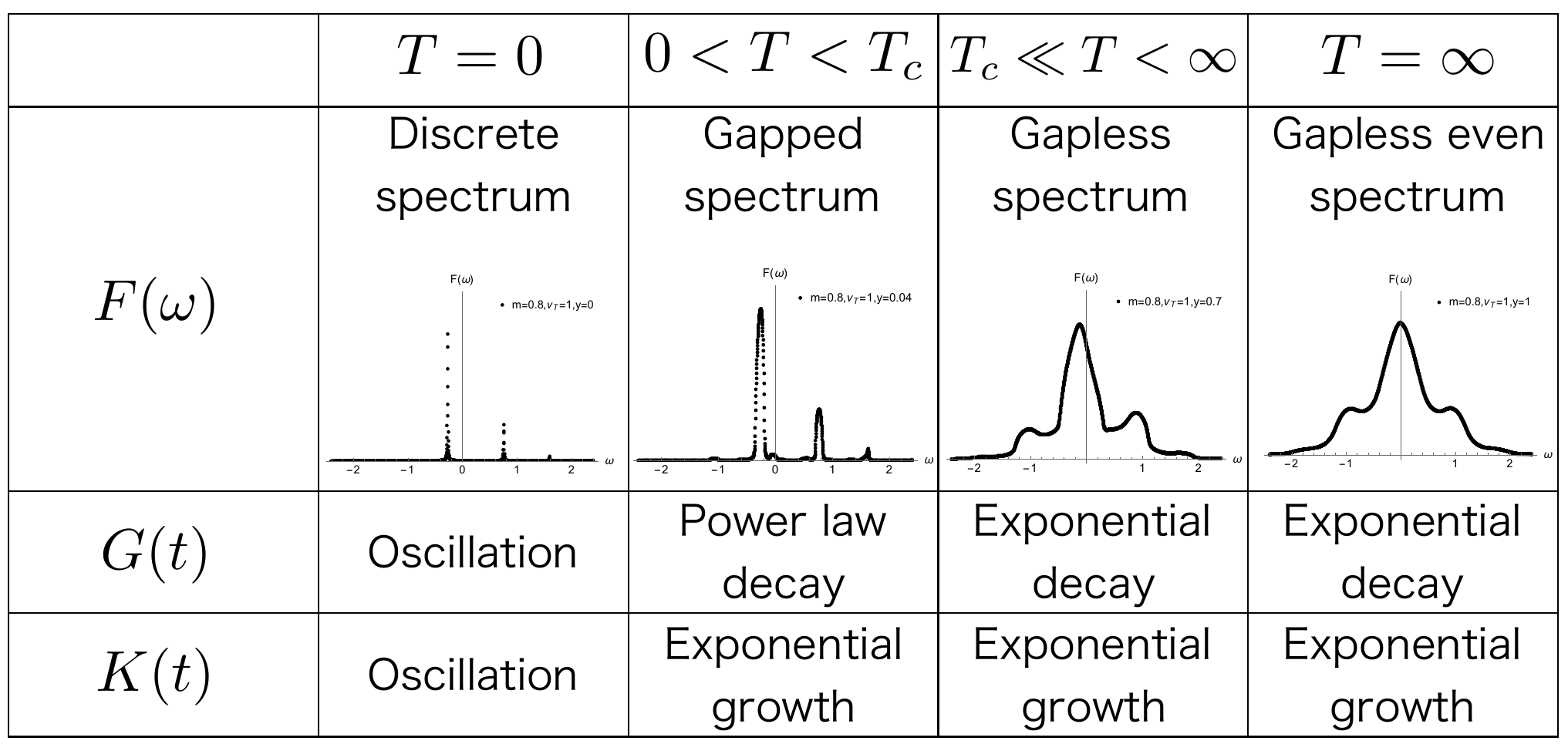}
       \caption{Summary table for the behaviors of the IP model. Here, $T_c$ is the critical temperature at which the spectrum becomes gapless. 
 The critical temperature in Figure \ref{fig:F(w)} is $y=e^{-m/T_c}\sim0.1$  for $m=0.2$, and $y=e^{-m/T_c}\sim0.3$  for $m=0.8$. The spectrum becomes smooth at high enough temperature $T_c\ll T$.}\label{fig:SummaryTable}
\end{figure}

In this paper, we study the Lanczos coefficients and Krylov complexity of the IP model in the temperature range between $T=0$ and $T = \infty$. To represent an infinite number of gaps in the spectrum of the IP model at nonzero low temperature, we consider a model consisting of infinite Wigner semicircles. Our analysis shows that the Lanczos coefficients $b_n$ show linear growth in $n$ with fluctuations at any nonzero low temperatures. Although the fluctuations of $b_n$ reduce the growth rate of the Krylov complexity, for {\it any} nonzero temperature $T > 0$, the Krylov complexity grows exponentially in time. This is due to the fact that the IP model spectral density consists of {\it infinite} cuts (with gaps) and asymptotically their amplitudes decay exponentially in $\omega$. 
We also study the Lanczos coefficients at high temperatures where the gap disappears but the spectrum has infinite local peaks, which are remnants of infinite cuts. In these temperatures, the Lanczos coefficients are linear in $n$, and there are almost no effects for $b_n$ by infinite peaks. Thus at high temperatures, the Krylov complexity also grows exponentially in time and we conclude that at any nonzero temperature, the Krylov complexity grows exponentially. However as the temperature becomes larger from zero, the slope of the exponential growth becomes larger. 
See Figure \ref{fig:SummaryTable} that summarizes the behaviors of the IP matrix model.

We also study the Lanczos coefficients of the IOP matrix model, a cousin of the IP matrix model, where the interactions preserve the number of adjoints. The resultant Lanczos coefficients are constants in the large $N$ planar limit, and those constants are determined by the 't Hooft coupling and temperature-to-mass ratio. Thus in the IOP matrix model, in the planar limit, the Krylov complexity grows only linearly in time even in the infinite temperature limit.  
We also study the $1/N^2$ corrections of the IOP matrix model.

Our conclusion is that in the IP matrix model, at any nonzero temperatures $T> 0$, the Krylov complexity grows exponentially in time. However in the IOP matrix model, at any temperature, the Krylov complexity never grows exponentially in time. Since the IP model at nonzero temperature reflects the nature of the deconfinement phase in the gauge theory, the result that the Krylov complexity grows exponentially at all nonzero $T > 0$ suggests that the Krylov complexity can play the role of the order parameter for the confinement/deconfinement in gauge theories. The reader might wonder if this is so given the fact that in quantum field theory, even in free theory limits, the Krylov complexity grows exponentially \cite{Dymarsky:2021bjq, Avdoshkin:2022xuw, Camargo:2022rnt}.  This is because in free theory for noncompact space, the spectrum becomes continuous due to the continuity of the momentum. However, if we put the gauge theory on the compact space, for example, ${\cal{N}}=4$ SYM on $S^3$, then the spectrum is discrete at the confinement phase and continuous in the deconfinement phase in the large  $N$ limit. As we have seen, Krylov complexity grows exponentially if the spectrum is continuous, has no upper bound, and decays exponentially. Therefore in such settings, we conjecture that the Krylov complexity plays the role of the order parameters for confinement/deconfinement phase transitions, not only as examined in \cite{Avdoshkin:2022xuw, Kundu:2023hbk}, but also in general settings.  We will report on the analysis of the Krylov complexity for large $N$ gauge theories in \cite{AnegawaIizukaNishida}.

It would be better to understand more the relations between the out-of-time ordered correlators (four-point function) and the Krylov complexity. 
One obvious and missing calculation is the OTOCs calculation in the IP model for $m > 0$ at nonzero temperature. By the comparison between the OTOC time dependence and Krylov complexity's time dependence, we would like to understand better for black hole physics from dual gauge theories.


\acknowledgments
The work of NI was supported in part by JSPS KAKENHI Grant Number 18K03619 and also by MEXT KAKENHI Grant-in-Aid for Transformative Research Areas A ``Extreme Universe'' No.~21H05184. M.N.~was supported by the Basic Science Research Program through the National Research Foundation of Korea (NRF) funded by the Ministry of Education (RS-2023-00245035). 

\appendix
\section{Lanczos coefficients and Krylov complexity}
\label{review}

\subsection{Lanczos coefficients}

Lanczos coefficients can be calculated as follows  
\cite{Parker:2018yvk, RecursionBook}. Let us consider a local operator $
\hat{\mathcal{O}}$. Its time evolution is given by the Baker-Campbell-Hausdorff formula
\begin{align}
\hat{\mathcal{O}}(t) =e^{iHt}\hat{\mathcal{O}}e^{-iHt} 
=e^{i\mathcal{L}t}\hat{\mathcal{O}}  \,, \quad \mbox{where}  \quad \mathcal{L} := [H, \, \cdot \, \, ] \,.
\end{align} 
Here $H$ is a local Hamiltonian, which is Hermitian. 
Thus the operator $\hat{\mathcal{O}}$ keeps spreading over the subspace of the Hilbert space and how quickly it spreads is our interest.

For canonical ensemble, we define the following inner product between operators $\hat{A}$ and $\hat{B}$
\begin{align}\label{innerproductfiniteT}
(\hat{A}\vert \hat{B})_\beta:=\frac{1}{Z}\text{Tr}[e^{-\beta H}\hat{A}^\dagger \hat{B}], 
\;\;\;  Z:=\text{Tr}[e^{-\beta H}],
\end{align}
where $\beta$ is the inverse temperature and Tr is over all $H$ eigenstates. One can define and check for any $n$
\begin{align}
(\hat{A}\vert\mathcal{L}^n\vert \hat{B})_\beta
& :=(\hat{A}\vert \mathcal{L}^n\hat{B})_\beta=(\mathcal{L}^n\hat{A}\vert \hat{B})_\beta.
\end{align}
Then we can construct the Krylov basis that follows 
\begin{align}
\label{orthonormalbasis}
(\hat{\mathcal{O}}_m\vert \hat{\mathcal{O}}_n)_\beta=\delta_{mn} \quad \mbox{(orthonormal basis)}
\end{align} 
The Lanczos coefficients by operator form are 
\begin{align}
  \hat{\mathcal{O}}_{-1} &:= 0\,, \; \quad  \hat{\mathcal{O}}_0:=  \hat{\mathcal{O}}, \; \\
\mathcal{L}\hat{\mathcal{O}}_n 
&= a_n\hat{\mathcal{O}}_n+b_n\hat{\mathcal{O}}_{n-1}+b_{n+1}\hat{\mathcal{O}}_{n+1} \notag\\
&=\sum_{m=0}\hat{\mathcal{O}}_mL_{m,n} \;\;\;(n \ge0 ), \label{recursionOnfiniteT}
\end{align}
where $L_{m,n}$ is expressed by 
\begin{align}
L_{m,n}:= \, (\hat{\mathcal{O}}_m\vert \mathcal{L}\vert \hat{\mathcal{O}}_n)_\beta=&
\begin{pmatrix}
a_0&b_1&0&\cdots\\
b_1&a_1&b_2&\cdots\\
0&b_2&a_2&\cdots\\
\vdots&\vdots&\vdots&\ddots\\
\end{pmatrix}.\label{MatrixLanczosfiniteT}
\end{align}
which is Hermitian. 
By using (\ref{recursionOnfiniteT}), we also obtain
\begin{align}
(\hat{\mathcal{O}}_m\vert\mathcal{L}^k\vert\hat{\mathcal{O}}_n)_\beta=(L^k)_{mn}.\label{Lkmn}
\end{align}
The Lanczos coefficients can be calculated from a two-point function $C(t;\beta):=(\hat{\mathcal{O}}\vert\hat{\mathcal{O}}(-t))_\beta$ for a given operator $\hat{\mathcal{O}}$.  Here $\hat{\mathcal{O}}$ is a normalized operator such that $C(t;\beta)=1$. We can compute $a_n$ and $b_n$ for $\hat{\mathcal{O}}_0=\hat{\mathcal{O}}$ by the following moment method. Let us define moments $M_{n}$ by using the Taylor expansion coefficients of $C(t;\beta)$ at $t=0$:
\begin{align}
M_n &:=\frac{1}{(-i)^{n}}\frac{d^{n}C(t;\beta)}{dt^{n}}\Big\vert_{t=0}=(\hat{\mathcal{O}}_0\vert\mathcal{L}^n\vert\hat{\mathcal{O}}_0)_\beta.
\end{align}
One can also compute moments $M_{n}$ by using a Fourier transformation of $C(t;\beta)$:
\begin{align}
\hspace{-4mm}M_{n}=&\int_{-\infty}^{\infty}\frac{d\omega}{2\pi}\,\omega^{n}f(\omega),\;\,\,
 f(\omega):=\int_{-\infty}^{\infty}dt\,e^{i\omega t}C(t;\beta).\label{mufw}
\end{align}
Then, 
one obtain relations between the Moments and the Lanczos coefficients. For example,
\begin{align}
\hspace{-8mm}M_1=(\hat{\mathcal{O}}_0\vert\mathcal{L}\vert\hat{\mathcal{O}}_0)_\beta= a_0,\,\,
M_2=(\hat{\mathcal{O}}_0\vert\mathcal{L}^2\vert\hat{\mathcal{O}}_0)_\beta=a_0^2+b_1^2. 
\end{align}

Through eq.~(\ref{mufw}), the high-frequency behavior of $f(\omega)$ and the asymptotic behavior of $b_n$ at large $n$ are correlated \cite{Lubinsky:1988}. In classical systems, the exponential tail of $f(\omega)$ has been proposed as a probe of chaos \cite{Elsayed:2014chaos}. Thus a relationship between chaos and the behavior of $b_n$ is expected in quantum systems and this is the motivation behind \cite{Parker:2018yvk}.

\subsection{Krylov complexity}  

The Krylov complexity is defined as follows: 
by decomposing the operator $\hat{\mathcal{O}}(t)$ into orthonormal bases, 
\begin{align}
\hat{\mathcal{O}}(t): =\sum_{n=0}i^n\varphi_n(t)\hat{\mathcal{O}}_n \,,
\end{align}
$i^n \varphi_n(t)$ is defined as a coefficient of the orthonormal basis. From the orthonormality eq.~\eqref{orthonormalbasis}, we obtain 
$\varphi_n(t)=i^{-n}(\hat{\mathcal{O}}_n\vert\hat{\mathcal{O}}(t))_\beta$, 
where 
$\varphi_n(t)$ satisfies
\begin{align}
\frac{d\varphi_{n}(t)}{dt}=ia_n \varphi_{n}(t)-b_{n+1}\varphi_{n+1}(t) +b_{n}\varphi_{n-1}(t) \,,\label{recursionwf}
\end{align}
with $\varphi_{-1}(t):=0$, $\varphi_{0}(t)=C(-t;\beta)$. 
The Krylov complexity $K(t)$ was introduced in \cite{Parker:2018yvk}, as 
\begin{align}
\label{Krylovdef}
K(t):=\sum_{n=1}^\infty n\vert\varphi_n(t)\vert^2 \,,
\end{align}
and it is conjectured in \cite{Parker:2018yvk} that this $K(t)$ is a good diagnostic for operator growth in the Krylov basis.

\bibliography{Ref}
\bibliographystyle{JHEP}

\end{document}